\documentclass[%
 reprint,
superscriptaddress,
 amsmath,amssymb,
 aps,
]{revtex4-2}
\usepackage[normalem]{ulem}
\usepackage[usenames,dvipsnames]{color}
\usepackage{graphicx}
\usepackage{dcolumn}
\usepackage{bm}

\usepackage[export]{adjustbox}

\begin{document}
 
\preprint{APS/123-QED}

\title{Effect of Exchange Interaction and Spin-Orbit Coupling on Spin Splitting in CdSnX (X = S, Se and Te) nanoribbons}
\author{Sutapa Chattopadhyay}
 \affiliation{Department of Physics, Savitribai Phule Pune University, Pune 411 007, India}
\author{Vikas Kashid}%
 \affiliation{Department of Physics, Savitribai Phule Pune University, Pune 411 007, India}%
 \affiliation{MIE-SPPU Institute of Higher Education, Doha -- Qatar}
\author{P. Durganandini}
 \affiliation{Department of Physics, Savitribai Phule Pune University, Pune 411 007, India}
\author{Anjali Kshirsagar}
\affiliation{Department of Physics, Savitribai Phule Pune University, Pune 411 007, India}

\date{\today}

\begin{abstract}
We have studied the topological properties of free standing Sn doped cadmium chalcogenide (CdSnX,  X = S, Se and Te) nanoribbons of varying widths and different edge structures.  The CdSnX nanoribbons are derived from free-standing hexagonal buckled monolayers, with three types of edges viz., distorted armchair, normal armchair and normal zigzag edges.  The unsatisfied bonds of X and Sn atoms at the edges cause non-zero values of the magnetic moment. This introduces an exchange field leading to inverted band structure.  
  The electronic band structures of 
 distorted armchair edge  nanoribbons also 
 exhibit different types of spin splitting property
 for different X atoms due to the different local orbital
 angular momentum at specific X atomic site with the inclusion of spin orbit coupling (SOC).   
  The gap opening at the band crossings near the Fermi level after inclusion of SOC are mainly due to SOC of Sn atom and are responsible for the electron and hole pockets making the system topologically exotic.  All the distorted edge nanoribbons show metallic behaviour with non-zero magnetic moments. Amongst CdX (X = S, Se and Te) nanoribbons, systems containing S atoms exhibit Weyl-like semi-metallic behavior and not much change with width, that of Se atoms  exhibit Zeeman-type spin splitting and significant change with varying width, whereas systems containing Te atoms show signature of Rashba spin splitting along with Zeeman-type spin splitting and moderate change with varying width.  The armchair edge nanoribbons show wide gap semiconducting behaviour.
 Zeeman-type  spin splitting is seen in the valence band region for systems containing S atoms and 
 Rashba spin splitting is visible in the conduction band region
  for systems containing  Se and Te atoms.  For  zigzag edge nanoribbons, no such signature of spin splitting is observed although all the nanoribbons acquire very high magnetic moments.
\end{abstract}

\maketitle

\section{Introduction}
Metal chalcogenides form a  class of novel semiconductors being studied extensively in last few years.  Among the metal chalcogenides, cadmium 
chalcogenides is the most common and useful family.  Cadmium chalcogenides CdX (X = S, Se and Te) are known to 
crystallize in both cubic zinc blende (ZB) and hexagonal  wurzite (WZ) structures. 
In the  bulk form, they have   wide and direct band gap leading to their unique 
semiconducting, electrical, physical and optical
properties. Due to the advances in nanoscale technology, the 
study of lower dimensional structures of cadmium chalcogenides  has  gained 
importance~\cite{Vasiliev2018,Khan2020}.

CdSe monolayer  having WZ phase was first prepared on Au (111) surface by electrochemical   atomic layer epitaxy~\cite{lister}. Few reports of synthesis of free-standing single layered lamellar-structured CdSe nanosheets showing
growth along [000$\bar{1}$] and [1$\bar{1}$00] axes of hexagonal WZ structure~\cite{son}, Sn doped CdSe thin films with X-ray diffraction peaks of (100) plane of WZ structure~\cite{kaur}, tin-incorporated nanocrystalline CdSe thin films along cubic (111) plane~\cite{sahu} and thin films of Sn doped CdS on glass substrate~\cite{das} are also
available in the literature.
It is well known that ZB (111) plane and WZ (100) plane have the same arrangement of atoms except that the 
stacking of layers is different in the bulk. Therefore, a two-dimensional (2D) structure 
CdX (X = S, Se and Te) is most likely similar to the cubic ZB (111) plane which is a honeycomb structure.   Zeng \textit{et al. } have theoretically studied electronic structure of 32 stable 2D 
honeycomb monolayer structures of II-VI semiconductors  using density functional theory (DFT)~\cite{zheng}. 
Among these HgTe shows topological nontrivial phase under appropriate in-plane tensile strain and on inclusion of spin-orbit coupling (SOC).

There is lot of interest  in low-dimensional structures of CdX from the point of view of nontrivial topological properties. SOC can also initiate topological character in a 2D structure through band inversion without application of an magnetic field~\cite{kane1}. 
 We have examined the electronic structure of Sn doped CdX monolayers; however,  in spite of band inversion deep in the
valence and conduction bands between Sn $s$ and $p$ and X $p$ states, these monolayers lack topological properties~\cite{sutapa}.

Besides thin atomic films exhibiting topological phases,  the 2D structures when confined laterally, resulting in quasi one-dimensional (1D) structures with specific edge structures are equally attractive.   Changing the edge style and  width offers versatility to tune various properties including topological properties of a material ~\cite{rama, maity}.  In this context,  graphene nanoribbons have been studied extensively for topological properties~\cite{ zhao2021, wang2021, rizzo2018}.


In the present paper, we have studied detailed electronic structure of a  new class of nanoribbons derived from Sn doped CdX (X = S, Se and Te) monolayers and have brought out the crucial role of SOC due to Sn in band inversion and of  SOC in the spin splitting as X changes from S to Se to Te. The formula unit for the Sn doped CdX nanoribbons is CdSnX$_2$, however,  we will refer to the nanoribbons as CdSnX nanoribbons. We  have considered  nanoribbons with 3 different types of edges viz., (i)~distorted armchair (henceforth referred to as distorted edge or d-CdSnX nanoribbons), (ii)~normal armchair (henceforth referred to as armchair edge or a-CdSnX nanoribbons) and (iii)~normal zigzag (henceforth referred to as zigzag edge or z-CdSnX nanoribbons).  Due to inversion symmetry breaking environment, different types of spin splitting are observed in zero magnetic field~\cite{Winkler}. At time-reversal invariant points and in the absence of space inversion symmetry,  we have seen  Zeeman-type and Rashba  spin-splitting prominently in d-CdSnTe nanoribbons with inclusion of SOC in the calculations.  The phenomenon, in which the spin degeneracy is lifted at the time invariant conjugate points due to SOC (in absence of applied magnetic field)  along the out of plane direction, is known as the Zeeman-type splitting. Rashba splitting, on the other hand, is lifting of the spin degeneracy along the in-plane direction in space symmetry broken system due to the application of SOC. In armchair edge nanoribbons, we see Zeeman-type and Rashba splitting but the nanoribbons turn into semiconductor. The Zigzag edge nanoribbons do not show much variation with width except
CdSnTe nanoribbons and all have non-zero cell magnetic moment with values larger than corresponding distorted edge CdSnX nanoribbons.

The structure of the paper is as follows: Section~\ref{methods} discusses the computational methodology used for calculating the electronic structure and topological properties and 
  Results of  detailed electronic structure and  topological properties are presented in 
Sec.~\ref{structure}. Section~\ref{Discussions}
presents the summary and conclusions of the work.

 \section{Computational Details}\label{methods}
\begin{figure*}[hbt!]
  {\includegraphics[width=6.8in]{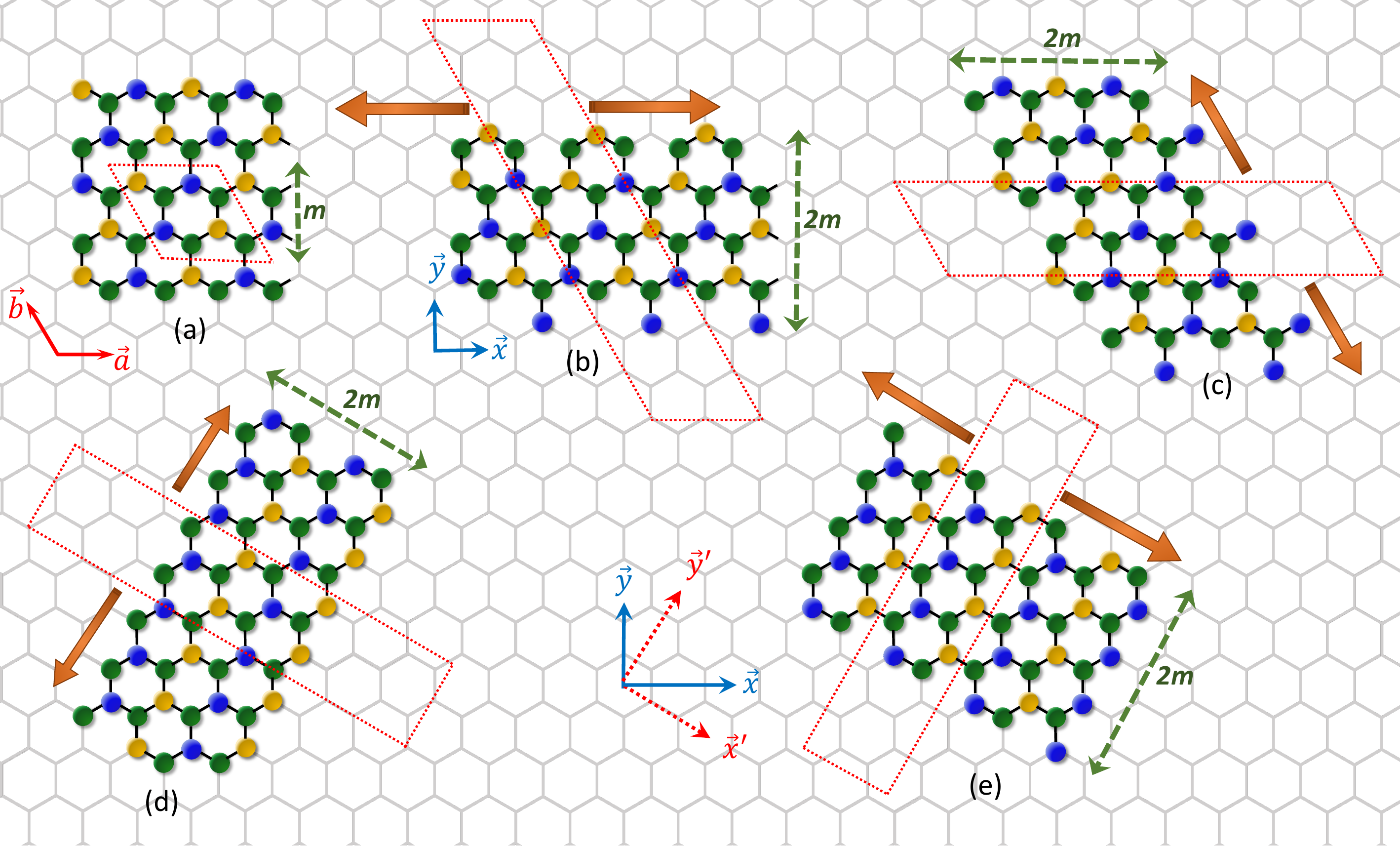}}    
\caption{(a) Optimized geometry of hexagonal Sn doped monolayer of CdX (X = S, Se and Te) in $xy$-plane. The green, yellow and blue balls
represent chalcogen (S, Se or Te), cadmium and tin atoms respectively. The lattice vectors  $\vec{a}$ and $\vec{b}$  for hexagonal lattice are shown by red coloured arrows with respect to the grey coloured uniform hexagonal sheet in the background. The Cartesian axes $\vec{x}$ and $\vec{y}$ are shown for reference. CdSnX nanoribbons shown in (b) and (c) are periodic along $\vec{a}$ and $\vec{b}$, respectively.  The nanoribbons in (b) and (c) are formed by cutting Sn doped CdX (X = S, Se and Te) monolayers (shown in (a)) in direction parallel to the lattice vectors and by introducing vacuum appropriately. These nanoribbons have distorted armchair edges. We then rotate the monolayer through an angle ($\pi /6$) radians in the clock-wise direction in the $xy$-plane about the $z$-axis so that the Cartesian axes $x$ and $y$ become the Cartesian axes $x'$ and $y'$ and choose the lattice vectors of the monolayer parallel to the Cartesian axes $x'$ and $y'$. The nanoribbons shown in (d) and (e) are obtained from this transformed monolayer by cutting the nanoribbons parallel to Cartesian axes 
$y'$ and $x'$ and introducing vacuum appropriately.  These nanoribbons exhibit normal zigzag and normal armchair edge topologies respectively. The dotted parallelograms  in (a), (b) and (c) and the dotted rectangles in (d) and (e) indicate the unit cells of the respective structures with sufficient vacuum in the non-periodic directions. The widths (2 in the figure) of the nanoribbons along the non-periodic direction are denoted in units of $m$ and are shown for (b)-(e) in the figure.  The arrows in orange colour indicate the direction of periodicity. }
\label{mon} 
\end{figure*}

We have performed calculations based on spin-polarized density functional theory (DFT),  using Vienna \textit{ab initio} simulation package (VASP)~\cite{VASP2,VASP3},  to compute the electronic properties of CdSnX  nanoribbons based on projected augmented wave method of
 Bl\"{o}chl~\cite{paw}.  Perdew-Burke-Ernzerhof (PBE) generalized gradient 
 approximation (GGA) is used for the exchange-correlation energy functional~\cite{pbe}. The kinetic energy cut off is set to  $500$~eV for all the calculations. The  convergence thresholds for energy and force are set to
 $10^{-6}$~eV and 0.001~eV/\AA\ respectively during the structural relaxation of Sn doped CdX (X = S, Se and Te) monolayers, constructed from CdX (111) surface of ZB  structure with two Cd atoms substituted by two Sn atoms. The monolayer unit cell is planar hexagon consisting of 2 Cd, 2 Sn and 
 4 X atoms as shown in Fig.~\ref{mon}(a).  More details about the monolayer results can be found in our previous work~\cite{sutapa}.   
 
   Sumo and PYPROCAR,  python based codes, are 
 used for band structure plots~\cite{sumo, procar}  and  VASPKIT code for spin density plots~\cite{vaspkit}.

 The relaxed structures of pristine monolayers are almost planar however upon Sn doping,  the monolayer structure exhibits buckling 
$\sim$~1.25~\AA~ along the $z$-direction. 
  CdSnX nanoribbons are constructed from the optimized Sn doped CdX monolayers such that the nanoribbons are infinite (periodic) either along one of the lattice vectors (as shown in Figs.~\ref{mon}(b) or ~\ref{mon}(c)) or along one of the   Cartesian axes (as shown in Figs.~\ref{mon}(d) or ~\ref{mon}(e)) by appropriately cutting the monolayer and introducing
  sufficient vacuum ($\geq$ 15~\AA) in the non-periodic directions to avoid interaction of periodic images.   The nanoribbons are neither relaxed nor passivated. Only self-consistent total energy calculations are carried out without and with inclusion of spin-orbit coupling (SOC). Figure~\ref{mon} illustrates the geometries of the various nanoribbons (Please refer to the supplementary information (SI) for more details). The unit cell vectors of 
nanoribbons, shown in Figs.~\ref{mon}(d) and \ref{mon}(e) are parallel to Cartesian axes  $x'$ 
and $y'$, whereas in case of distorted nanoribbons, (Fig.~\ref{mon}(b) and \ref{mon}(c)), the unit cell vectors are parallel to the lattice vectors of the hexagonal lattice. 

If the Cartesian $x-$axis and lattice vector $\vec{a}$ are aligned for the nanoribbon, then $y-$axis is at $30^{\circ}$ from the lattice vector $\vec{b}$ making these nanoribbons tilted.   Such kind of tilted nanoribbon structures have attracted much interest due to the edge magnetic moment and 
thermal stability~\cite{rama,subhan,farooq}

 The width  $W$ of the ribbon is quantified as $W = Nm$ where $N$ is 
 the number of the unit cells along the finite size and $m$ is the lattice constant (all nanoribbons of width 2m are shown in Fig.~\ref{mon}) of the monolayer unit cell.
    Electronic structure studies are carried out for widths 2$m$ to 5$m$.  

    All the distorted edge, armchair edge and zigzag edge nanoribbons show different atomic configuration 
 for the two terminating edges as explained earlier.  Hence, all the nanoribbons have broken space symmetry.  
 The details of the bond lengths and bong angles  are analogous to their values in respective monolayers and are listed in the Supporting information (SI) for completeness.

  \section{Results}\label{structure}

\subsection{Distorted edge Sn doped CdX Nanoribbons} 
The geometrical structure of distorted edge nanoribbons with lengths along lattice vectors $\vec{a}$
and $\vec{b}$ are equivalent.  Therefore, results of electronic structure calculations for the two are expected to be same. We have explicitly calculated the band structures, density of states (DOS),  partial DOS (PDOS) and spin density and have confirmed this to be indeed the case. Therefore, in this section, we 
discuss the electronic structure in the vicinity of the Fermi level (E$_F$) for nanoribbons shown in Fig.~\ref{mon}(c) only for  widths varying from 2$m$ to 5$m$.

Tilted nanoribbon structures seem to show unusual and interesting properties. We present a comparison of the electronic structures  to bring out the effect of inclusion of SOC and the effect of increasing width $W$, particularly on the specific bands in the vicinity of the Fermi energy.  We discuss further that the electronic properties of these nanoribbons are significantly determined by the edge structures of the nanoribbons and the type of the atoms on the edges~\cite{li,barone}.  The relative comparison is made based on the nature of the lowest conduction band (LCB) numbered as 1, the highest valence band (HVB) numbered as 2, and one band below the highest valence band (BVB) numbered as 3 for convenience although it is not customary to number the bands in this fashion.

    \subsubsection{d-CdSnS Nanoribbons}
\begin{figure}
{\includegraphics[scale=0.38]{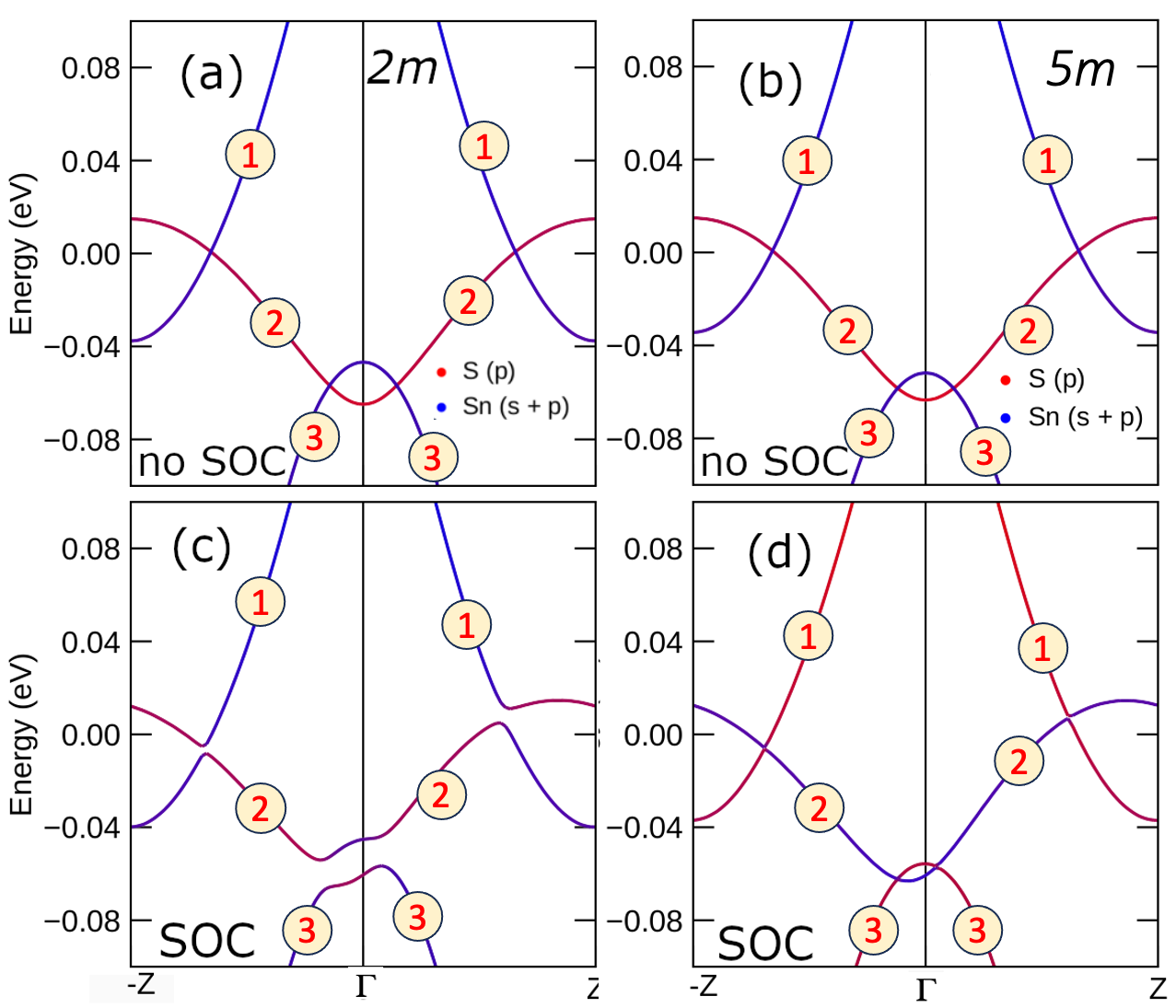}}  
 \caption{Band structure plots for d-CdSnS nanoribbons with respective Fermi energy set to zero.  Panels (a) and (c) respectively indicate the band structures without and with inclusion of SOC for the nanoribbons with  width 2$m$, whereas panels (b) and (d) respectively  show the band structures for the nanoribbon having  width 
 5$m$ without and with inclusion of SOC.  
 The red and blue curves represent relative contributions from S and Sn states respectively.}
 \label{CdSnS-band}
\end{figure}

The electronic band structures of CdSnS distorted edge nanoribbons of width 2$m$ and 5$m$ are shown in Figs.~\ref{CdSnS-band}(a) and \ref{CdSnS-band}(b)
without inclusion of SOC and in  Figs.~\ref{CdSnS-band}(c) and \ref{CdSnS-band}(d) with inclusion of SOC. The band structure without SOC shows band-crossing at two time-irreversible points at E$_F$.
Such kind of band crossing resembles Weyl band structure reported earlier~\cite{wely1,wely2,wely3}. 
Weyl semimetal state is the signature of broken inversion and time reversal symmetry 
which is present in this nanoribbon structure.
The valence (band number 2) and conduction (band number 1)  bands are inverted and 
their crossings at the Fermi level have the contribution  of 
spin-up and spin-down characters respectively as shown in 
Fig.~\ref{CdSnS-band}(a)  coming
  from the $p$ states of  Sn and S atoms
  from the same edge (right edge) of the nanoribbons. Such inverted bands appear due
to the exchange field and the system can be in a topologically
nontrivial
phase~\cite{Yu}. 
 
The majority spin bands near the Fermi energy mostly arise due to $p$-$p$ hybridization between Sn and S states,  whereas the minority spin bands are mostly due to $s$-$p$ hybridization.  These bands  contribute to band crossing just below the Fermi energy.
The $p$ states of Sn and S atoms split in
 $p_x$, $p_y$ and $p_z$  due to the presence of the crystal field.
 
\begin{figure}
{\includegraphics[scale=0.35]{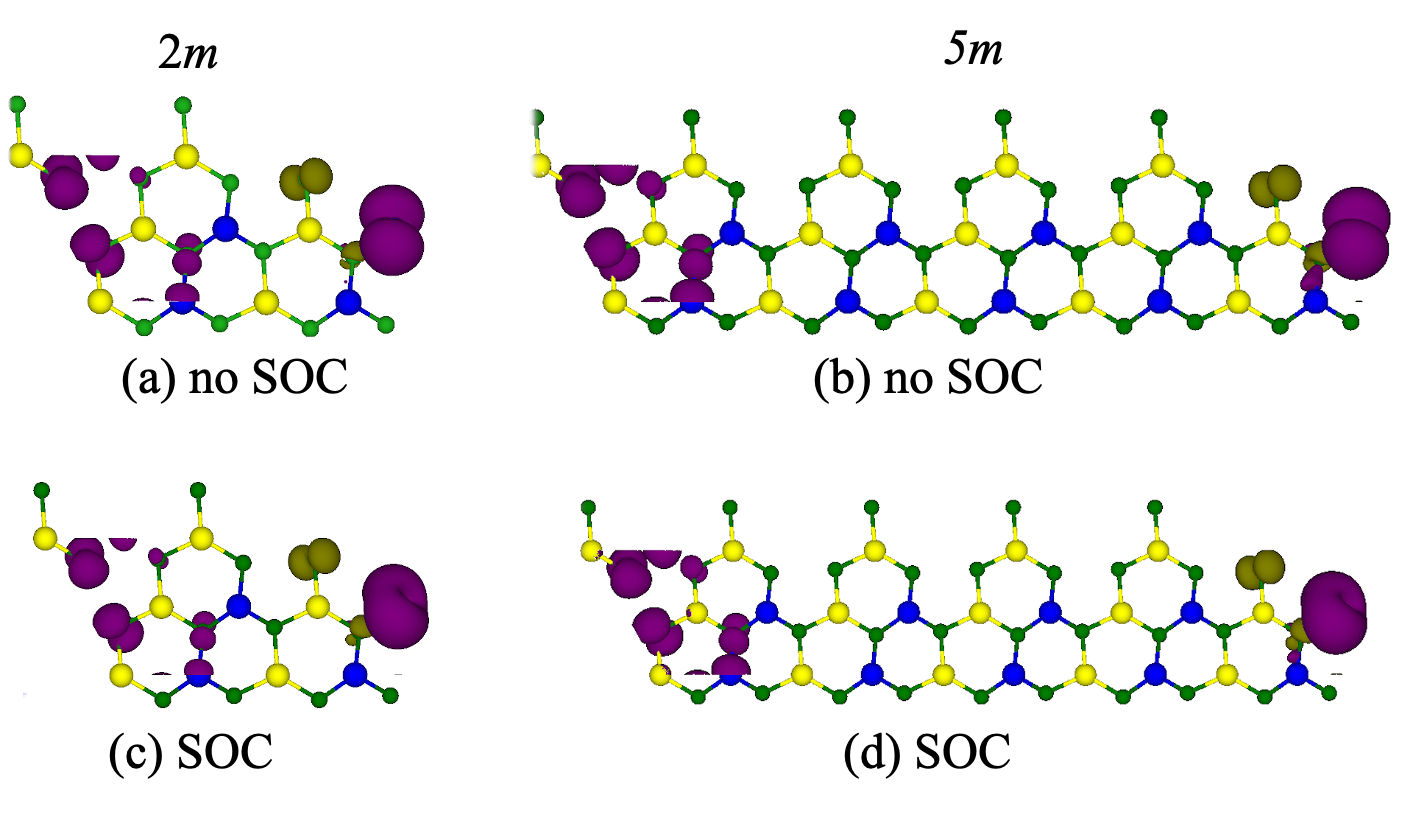}}  
 \caption{Spin density ($\rho_{\uparrow} - \rho_{\downarrow}$) distributions for d-CdSnS nanoribbons of widths 2$m$ and 5$m$ without inclusion of SOC are shown in (a) and (b) respectively. The spin density distributions for the same widths after inclusion of SOC are shown in (c) and (d) respectively.  The magenta and green isosurfaces indicate positive and negative values of spin densities respectively.}
 \label{CdSnS-spin}
\end{figure}

Figure~\ref{CdSnS-spin} depicts the spin density plots for widths 2$m$ and 5$m$ of CdSnS systems,  indicating that  only the edge Sn and S atoms gain magnetic moment.
Figures~\ref{CdSnS-spin}(a) and (b) are for calculations without SOC while Figs.~\ref{CdSnS-spin}(c) and (d)  are with inclusion of SOC for widths  2$m$ and 5$m$ respectively. As can be seen from the figures, the spin density results without and with inclusion of SOC are not much different.
Due to the reduced coordination of the edge Sn and S atoms,  these atoms exhibit magnetic moments, viz.,  Sn atoms at the right edge show a magnetic moment of  $0.43~\mu_{B}$, whereas S atoms on the opposite edge exhibit induced magnetic moment upto $0.18~\mu_{B}$ aligned parallel to the magnetic moment of the Sn atoms for both the widths 2$m$ and 5$m$. The S atoms bonded with the Sn atom on the right edge are aligned anti-parallel to the magnetic moment of Sn atoms with low values of the magnetic moment.  These non-zero magnetic moments introduce exchange field in the nanoribbons that affects the band splitting at E$_F$ (not visible in Figs.~\ref{CdSnS-band}(a) and \ref{CdSnS-band}(b)). 

Band inversion can lead to enhanced Van Vleck spin susceptibility in these structures~\cite{Yu}.  We have calculated band inversion strength
(BIS) by calculating  the difference  of the minimum of the conduction band 
and maximum of the valence band as discussed in Ref.~\cite{bahadur}.
BIS strength is used to quantify the nontrivial character of
the bands of the system~\cite{bahadur,Zhang}. Table~\ref{bis}
lists the BIS values and the momentum space separation
($\Delta k$) calculated without SOC.  The robustness of 
the topological property of the system is predicted with BIS 
and $\Delta k$ values.  Negative value of BIS  indicates that the 
system is topologically nontrivial while its positive value 
indicates that the system is susceptible to turn into gapped
insulator~\cite{bahadur}.

For all widths of d-CdSnS nanoribbons, with inclusion of SOC in the calculations, one prominent band crossing is observed approximately at 0.05 eV below Fermi energy between bands 2 and 3, as shown in Figs.~\ref{CdSnS-band}(c) and \ref{CdSnS-band}(d).  
 This band crossing takes place between up spin electrons of
right edge Sn atoms and  down spin $p_z$
states of left edge S atoms. 
The magnitude of SOC for S atoms is very negligible but 
Sn atoms introduce very strong SOC owing to their high atomic number (Z).
The band structure without SOC shows time reversal symmetry; however, after inclusion of SOC, the crystal field symmetry is broken which results into breaking of time reversal symmetry of the bands leading to the observation of band tilting for nanoribbons of all widths (Figs.~\ref{CdSnS-band}(c) and (d)). Inclusion of SOC opens up a gap for the tilted bands near the Fermi level (not really visible in Fig.~\ref{CdSnS-band}(d)) with the formation 
of electron and hole pockets.  The gap is a result of interaction of the atoms at the edges; hence as the
thickness of the CdSnS nanoibbons increases, the gap decreases. 
The magnitudes of band gap opening, due to 
SOC, are also listed in Table~\ref{bis} for all the widths of nanoribbons. The band gap values 
from Table~\ref{bis} also suggest that the CdSnS
3$m$ and 5$m$ show lower band gap values than those
for widths 2$m$ and 4$m$.  This change in band gap is due to
 the additional structural symmetry arising due to 
 odd and even values of $m$ for the nanoribbons~\cite{Magda,li1,yue}.
 The band  crossing between bands numbered 2 and 3  in  the valence region shows gap with $p$-$p$ band
inversion. 
As the width of the nanoribbons increases, the BIS value decreases
while value of $\Delta k$ increases but the semi-metallic property 
is consistently observed for calculations both without and with SOC.
Thus, the  broken symmetry at the edges, lead to strong crystal field and exchange field and band inversion in CdSnS nanoribbons of all widths.

\begin{table}
\caption{Band inversion strength (BIS)
in eV and $\vec{k}$-space separation
($\Delta k$) in {\AA}$^{-1}$ without the inclusion of SOC and the band gap opening in meV at -K and K points for calculations with inclusion of SOC for d-CdSnS nanoribbons. }
\setlength{\tabcolsep}{12pt}
\renewcommand{\arraystretch}{1.6}
\begin{tabular}{c c c c c}
\hline\hline
 Width & BIS  &$\Delta k$  & 
 \multicolumn{2}{c}{Band gap} \\
  & (eV) & ({\AA}$^{-1}$)&  $-K$ & $K$\\
  \hline\hline
 2$m$ & -0.248 & 0.585 & 3.70 & 7.50 \\

 3$m$ & -0.246 & 0.596 & 0.57 & 1.17\\
 
 4$m$ & -0.245 & 0.599 & 1.91 & 2.21 \\
 
 5$m$ & -0.240 & 0.593 & 0.59 & 1.18\\
 \hline
\end{tabular}
 \label{bis}
 \end{table}


    {\subsubsection{d-CdSnSe Nanoribbons}

     \begin{figure*}
 {\includegraphics[scale=0.53]{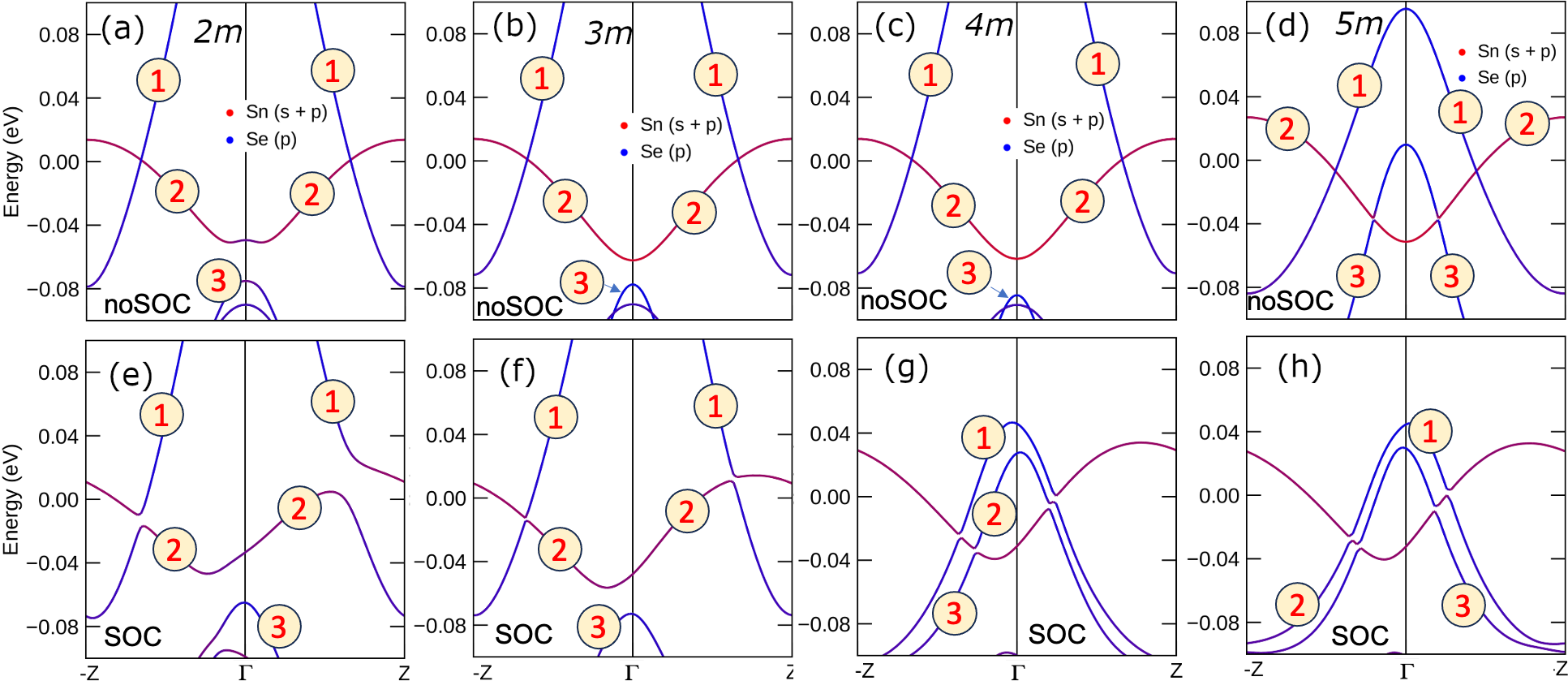}}  
 \caption{Band structure plots for d-CdSnSe nanoribbons with respective Fermi energy set to zero.  Panels (a)-(d) indicate the band structures for the nanoribbons of widths 2$m$ - 5$m$ respectively without inclusion of SOC, whereas panels (e)-(h) show the band structures  with inclusion of SOC.   
 The red and blue curves in (a)-(d) indicate  relative contributions from Sn and Se states respectively.}
 \label{CdSnSe-band}
\end{figure*}

    Figures~\ref{CdSnSe-band}(a) to (d) depict the band structure plots without the application of SOC for d-CdSnSe nanoribbons of widths varying from 2$m$ to 5$m$. As opposed to d-CdSnS nanoribbons, the band structures change a lot with width for d-CdSnSe nanoribbons. Therefore the plots are depicted for all widths. Those bands in the vicinity of the Fermi energy show strong hybridization of Se $p_x$, $p_y$ and $p_z$ states with Sn $p_x$ and $p_z$ states as revealed in the PDOS plots (Fig.~S2(a) to Fig.~S2(d)).
 With the change in the atoms in  the nanoribbons from S to Se,  the general 
 features of the  band structure around the Fermi energy, as regards the nature of the bands, remain  similar.
 We have numbered the three bands near the Fermi region and will discuss
 the changes in these bands only, for all the widths from 2$m$ to 5$m$. 
 All the nanoribbons shows Weyl semi-metallic character without the
application of SOC as the two bands 1 and 2 cross the Fermi level at two time- irreversible points.
 These two bands are mainly from the $p$ states of Sn and Se atoms from the opposite edges of the nanoribbons. Such tunnelling effect is observed earlier in Bi$_2$Se$_3$ thick slab~\cite{Yu}. 
  Bands 2 and 3 do not cross in CdSnSe nanoribbons, as in CdSnS nanoribbons for widths 2$m$ to 4$m$. The nanoribbon of width 2$m$ exhibits band inversion character at 
the $\Gamma$ point  between the bands numbered 2 and 3 as shown in Fig.~\ref{CdSnSe-band}(a).
 For nanoribbons of widths  3$m$ and 4$m$,  not much change is observed in the bands, however suddenly for nanoribbons of width 5$m$, the relative positions of the three bands change; band 1 is lowered in energy and bands 2 and 3 cross, in addition to the crossing of bands 1 and 2. Table~\ref{zee}
lists the BIS values  and the momentum space separation
($\Delta k$) along with the spin splitting for the nanoribbon with width 5$m$ calculated without inclusion of SOC. 
The spin density plots for calculations without SOC are shown in Figs.~\ref{CdSnSe-Spin}(a), (c), (e) and (g), for the four widths. The general feature are similar to the d-CdSnS nanoribbons.

Inclusion of SOC in d-CdSnSe nanoribbons show more prominent effects as compared to d-CdSnS nanoribbons due to comparatively large atomic number of Se than S atoms.  For nanoribbons of widths $2m$ and $3m$,  gap opens up  at the band crossings near Fermi energy.  As the width is increased to $4m$ and $5m$, the band structure shows dramatic changes as seen in Fig. \ref{CdSnSe-band}(g) and \ref{CdSnSe-band}(h) clearly indicating a strong Zeeman-type splitting.  The spin component resolved band structure after inclusion of SOC is shown in Fig. \ref{CdSnSe-spin-split} for nanoribbon of width 5$m$.   These bands also show spin flip across its spread along $\Gamma$-$Z$ direction. These results clearly indicate a strong Zeeman type splitting in d-CdSnSe nanoribbons.  Fig. \ref{CdSnSe-spin-split} shows  $S_{y}$ and $S_{z}$ spin components contributing significantly at the $\Gamma$ symmetry line.

The spin density plots, Figs.~\ref{CdSnSe-Spin}(b), \ref{CdSnSe-Spin}(d), \ref{CdSnSe-Spin}(f) and \ref{CdSnSe-Spin}(h), depict the change in the edge spin density for nanoribbons of width  2$m$ to 5$m$ and also reveal that only the Sn atom terminated edge
 contributes to the non-zero ( positive as well as negative)  spin density with the application of SOC.  The
 Se atom bonded with Sn atom on the right  edge contributes to the magnetic moment of the cell
  due to strong SOC while  the  Se atom bonded with Cd
 atom on the left edge does not contribute any magnetic moment from widths 4$m$ and 5$m$.
 

  \begin{figure*}
 {\includegraphics[scale=0.58]{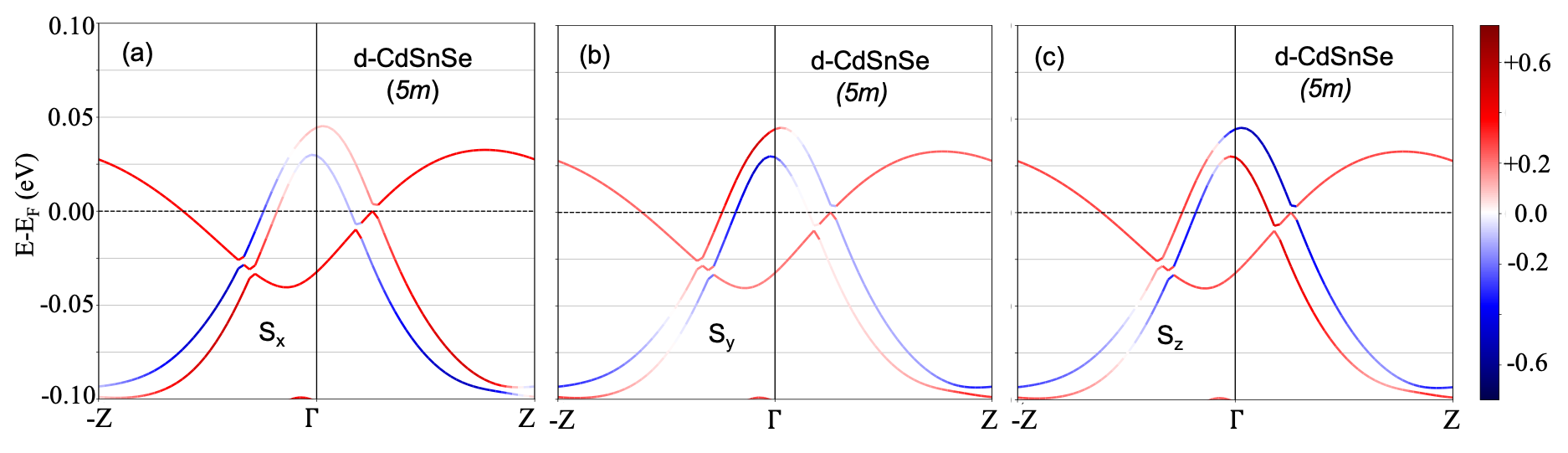}}  
 \caption{The band structure plots near the Fermi energy for d-CdSnSe nanoribbons for $5m$ width. Colors quantify
 the expectation values of  $S_{x}$, $S_{y}$ and $S_{z}$ spin components for the bands, after inclusion of SOC as shown in (a), (b) and (c) respectively. Red and blue indicate up and down directions respectively.}
  \label{CdSnSe-spin-split}
\end{figure*}


  \begin{figure}
 {\includegraphics[scale=0.44]{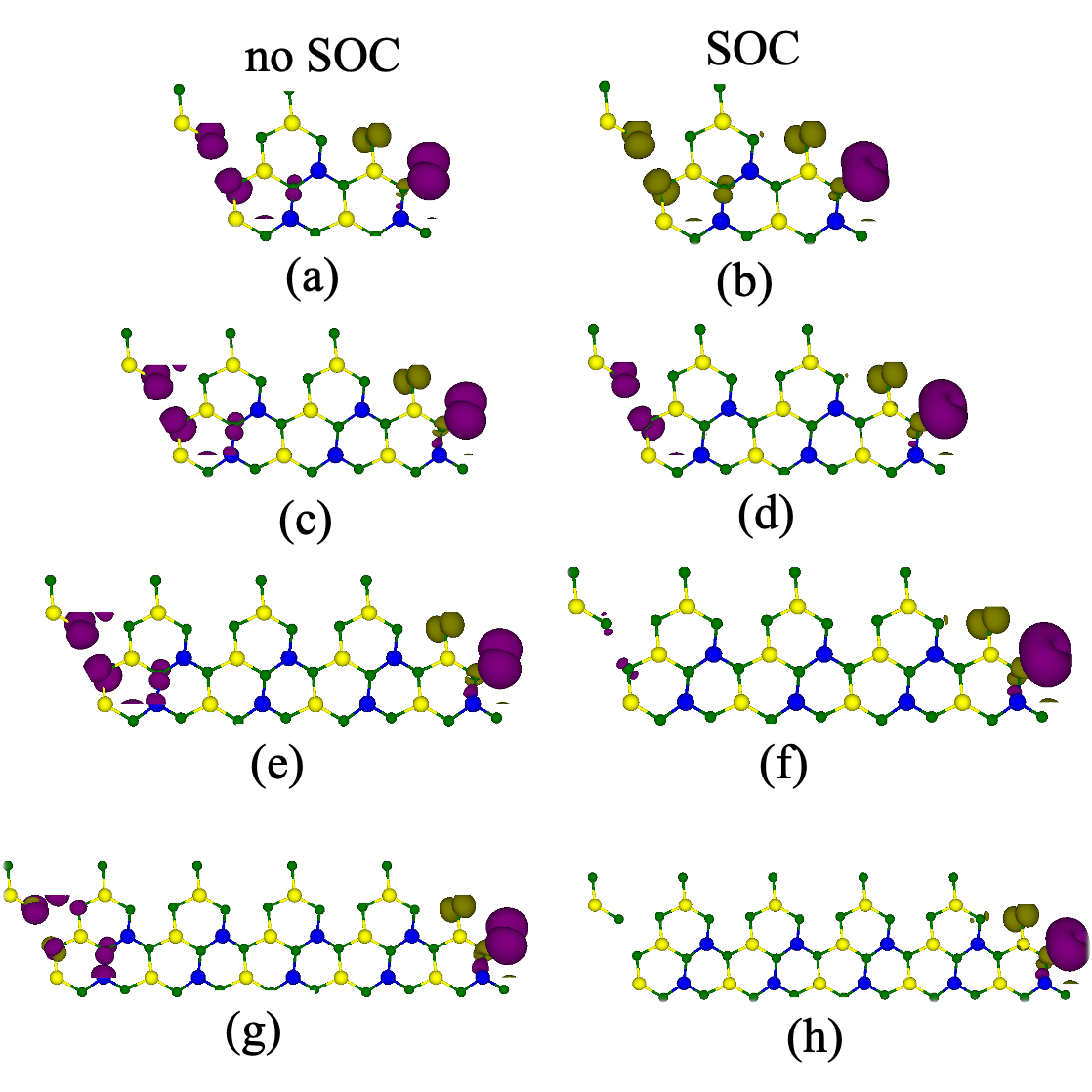}}  
 \caption{Spin density ($\rho_{\uparrow} - \rho_{\downarrow}$) distributions for d-CdSnSe nanoribbons of widths 2$m$-5$m$ without inclusion of SOC  are shown in the left panel  ((a), (c), (e) and (g) respectively)  and for the same widths after inclusion of SOC are  shown in the right panel ((b), (d), (f) and (h) respectively). The magenta and green isosurfaces indicate positive and negative values of spin densities respectively.}
  \label{CdSnSe-Spin}
\end{figure}


 \begin{table}
  \caption{{Band inversion strength (BIS) in eV for nanoribbons with widths
 2$m$ to 4$m$ and spin splitting (SS)  in eV for nanoribbon with width 5$m$
 and $\Delta k$ in {\AA}$^{-1}$ for d-CdSnSe nanoribbons without SOC.}}
\setlength{\tabcolsep}{23pt}
\renewcommand{\arraystretch}{1.6}
 \begin{tabular}{c c c }
 \hline\hline
  Width & BIS or SS  & $\Delta k$  \\
    &  (eV) & ($\AA{}^{-1}$) \\
 \hline\hline
  2$m$ & $-$0.3107 &0.570\\
 
  3$m$ & $-$0.3200 & 0.576\\
 
  4$m$ & $-$0.3180 & 0.573\\
 
  5$m$ & 0.0850 & 0.383\\
  \hline  
 \end{tabular}
 \label{zee}
 \end{table}

\subsubsection{d-CdSnTe Nanoribbons}
  
The electronic band structures for d-CdSnTe nanoribbons  are shown in Figs.~\ref{CdSnTe-band}(a) to \ref{CdSnTe-band}(d) for nanoribbons of widths 2$m$ and 5$m$ without and with inclusion of SOC. The  band inversion occurs in these nanoribbons even without inclusion of SOC. This is clearly visible in the band structure for nanoribbon of width 2$m$ around the $\Gamma$ point in Fig.~\ref{CdSnTe-band}(a) but not so obvious for nanoribbon of width 5$m$ in Fig.~\ref{CdSnTe-band}(b) since the small gaps between almost degenerate bands 1 and 2 and band 3, around -0.02~eV, are not visible. The states involved in this band inversion are mainly hybridized  Te $p$ and Sn $p$ states as revealed from Fig. S3 in SI.  The band inversion occurs between the up spin Sn $p$ states and down spin Te $p$ states.  
The conduction band is mainly formed by Sn $p$ and Te $p$ states.  
Figures~\ref{CdSnTe-band}(a) and \ref{CdSnTe-band}(b)
show the valence band crossing the Fermi level on both sides of the 
$\Gamma$ point.  Without inclusion of SOC, the band structure shows time reversal symmetry with the bands more degenerate for larger widths of the nanoribbon. Figure~\ref{CdSnTe-Spin} depicts the spin density distributions for d-CdSnTe nanoribbons of widths 2$m$ and 5$m$ without and with inclusion of SOC. The same edge Sn and Te atoms show magnetic moment of 0.39~$\mu B$ and 0.13~$\mu B$ with antiparallel alignment for nanoribbon of width 2$m$ and that of  0.4~$\mu B$ and 0.13~$\mu B$ again with antiparallel alignment for width 5$m$. These values of magnetic moments are smaller than those observed for distorted edge CdSnS and CdSnSe nanoribbons. No magnetic moment or spin density is observed at the other edge, unlike the d-CdSnS and d-CdSnSe nanoribbons. This may be a result of more covalent nature of bonds in CdSnTe nanoribbons.

\begin{figure}
 {\includegraphics[scale=0.36]{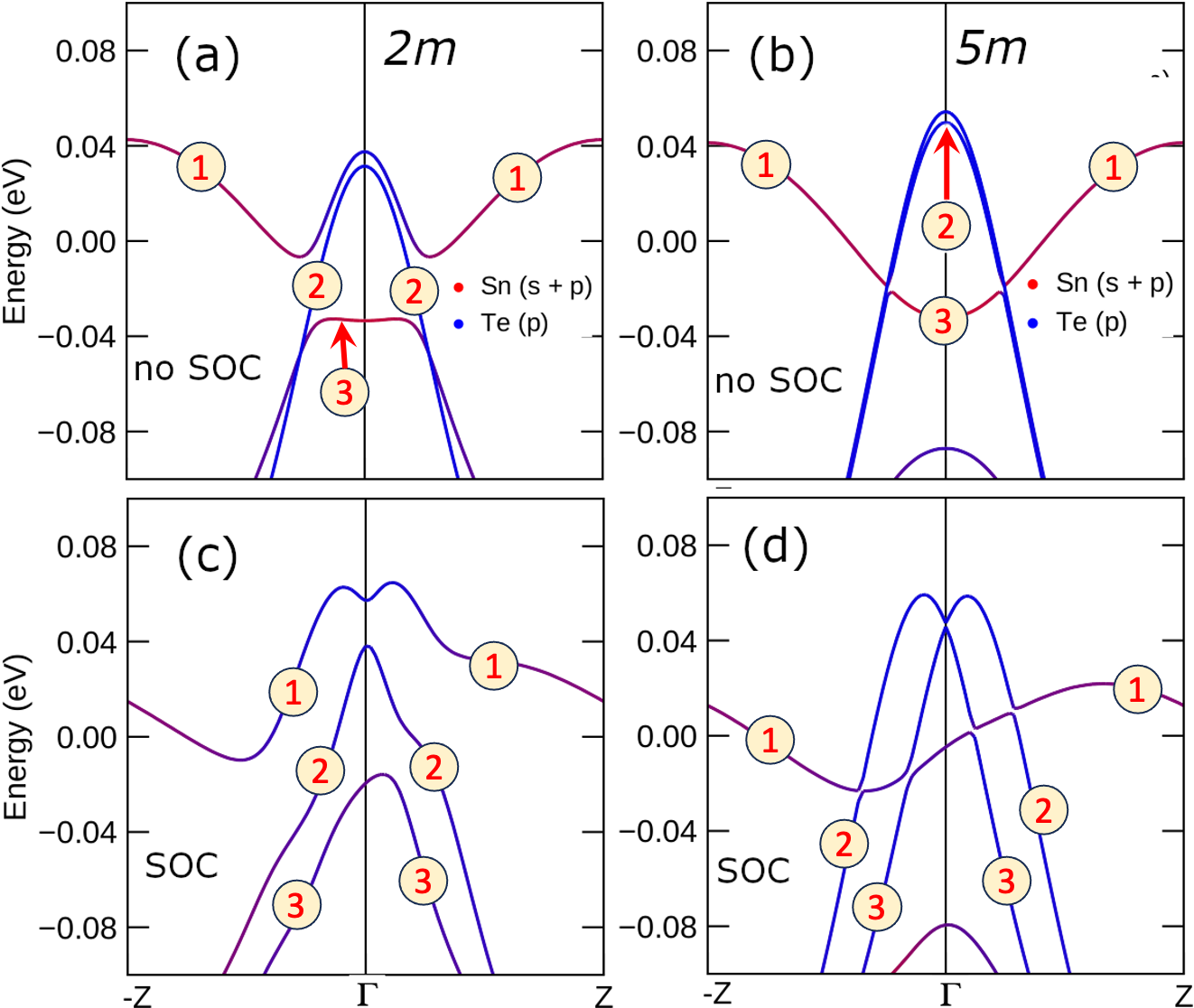}}  
 \caption{Band structure plots for d-CdSnTe nanoribbons with respective Fermi energy set to zero. Panels (a) and (c) indicate the band structures for the nanoribbons with width 2$m$ without and with inclusion of SOC, whereas panels (b) and (d) show the same for the nanoribbons with width  5$m$.  In (a) and (b), the red and blue curves represent relative contributions of Sn and Te states respectively.}
 \label{CdSnTe-band}
\end{figure}

\begin{figure}
{\includegraphics[scale=0.32]{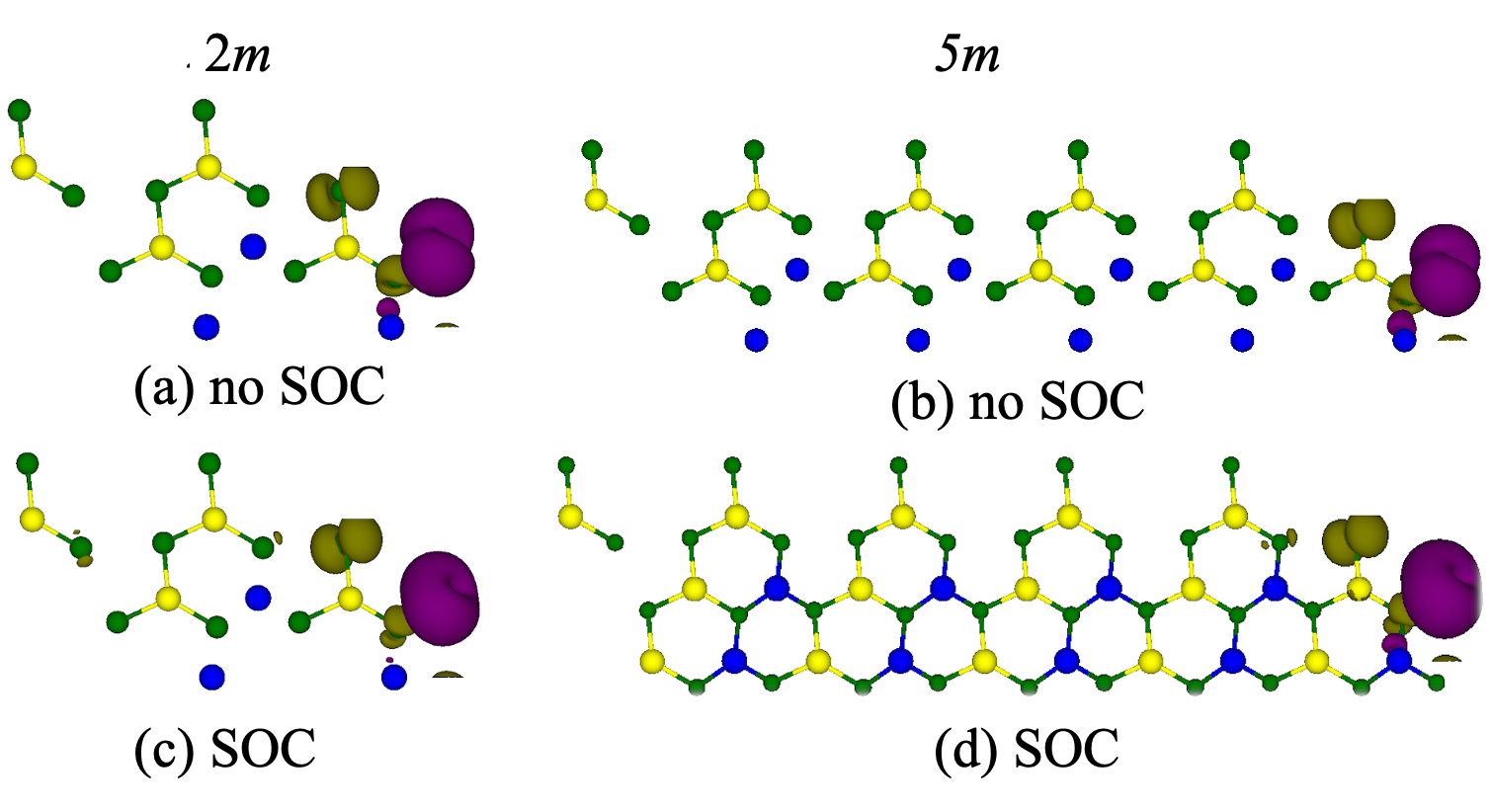}}  
 \caption{Spin density ($\rho_{\uparrow} - \rho_{\downarrow}$) distributions for d-CdSnTe nanoribbons of widths 2$m$ and 5$m$ without inclusion of SOC are shown in (a) and (b) respectively and after inclusion of SOC are shown in (c) and (d) respectively. The magenta and green isosurfaces indicate positive and negative values of spin density respectively.}
 \label{CdSnTe-Spin}
\end{figure}

  \begin{figure*}
 {\includegraphics[scale=0.58]{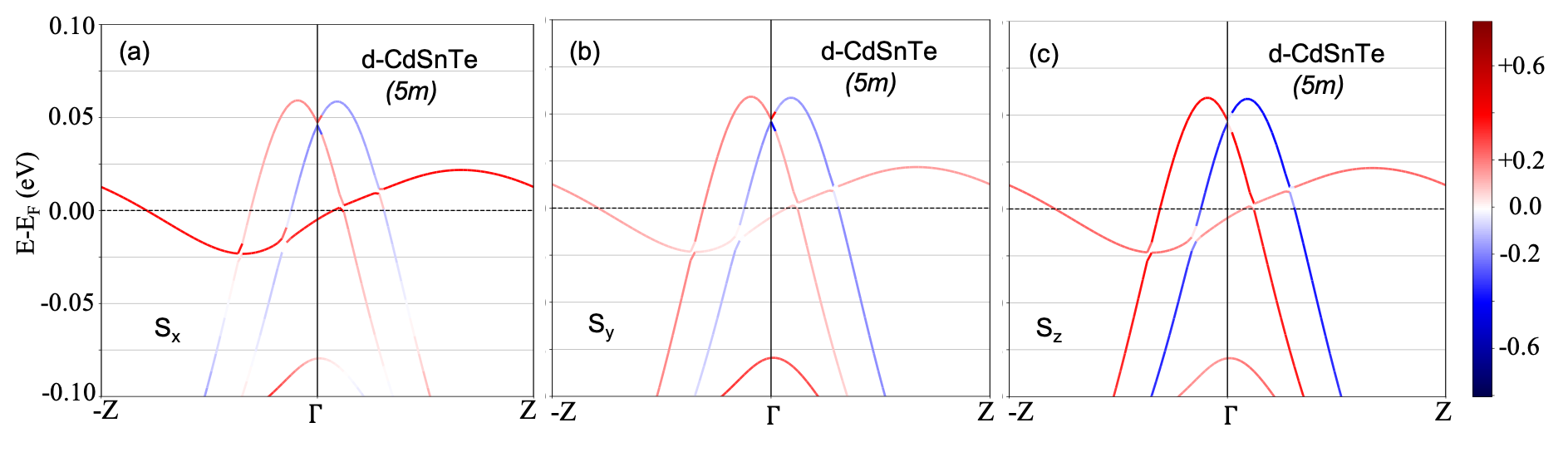}}  
 \caption{The band structure plots near the Fermi energy for d-CdSnTe nanoribbons for $5m$ width. Colors quantify the expectation values of  $S_{x}$, $S_{y}$ and $S_{z}$ spin components for the bands, after inclusion of SOC as shown in (a), (b) and (c) respectively. Red and blue indicate up and down directions respectively.}
  \label{CdSnTe-spin-split}
\end{figure*}

\begin{table}[h!]
\setlength{\tabcolsep}{5pt}
\renewcommand{\arraystretch}{1.6}
\caption{Spin splitting in meV,
$\Delta E$ in meV,  $\Delta k$ in {\AA}$^{-1}$  and  Rashba parameters  ($\alpha_R$) in  eV \AA{}$^{-1}$ with inclusion of SOC for d-CdSnTe nanoribbons.}
\begin{tabular}{c c  c c c}
 \hline\hline
 Width& Spin splitting   & $\Delta E$ & $\Delta k$  & $\alpha_R$\\
  &  (meV) & (meV) & (\AA{} $^{-1}$) & (eV\AA{}$^{-1}$)\\
 \hline\hline
 2$m$ & 6.4 & 7.6 & 0.042& 0.36\\
 
 3$m$ & 8.4&10.0 & 0.038 &0.526\\
 
 4$m$ & 7.6&9.3& 0.040 &0.465\\
 
 5$m$ & 4.7& 11.0& 0.041& 0.536\\
 \hline
\end{tabular}
\label{rashba}
 \end{table}

 \begin{figure*}
 {\includegraphics[scale=0.53]{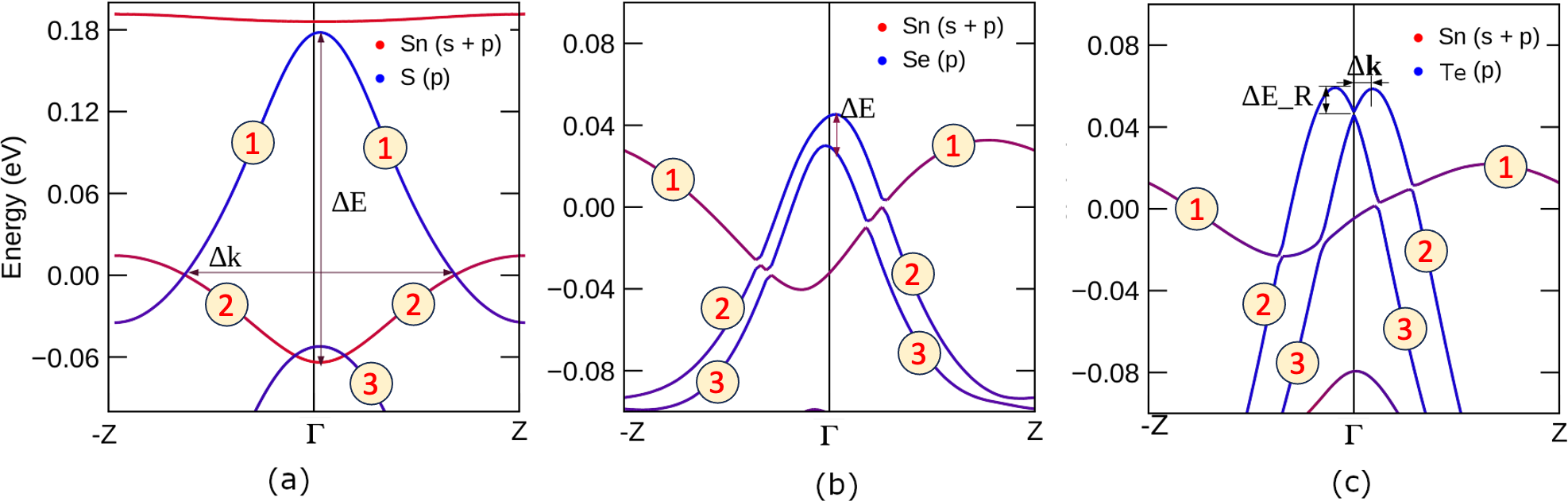}}  
   \caption{Variation in the type of spin splitting according to chalcogenide atoms present in the nanoribbons. (a)~Only band inversion is seen with the formation of Weyl points for  X = S nanoribbons, (b)~Zeeman-type spin splitting is observed in X = Se nanoribbons and (c)~Rashba spin splitting along with Zeeman-type splitting is observed in X = Te nanoribbons.}
  \label{spin}
  \end{figure*} 
Inclusion of  SOC induces band tilting leading to the Rashba spin splitting (RSS) of the conduction band
at $\Gamma$ point as depicted in  Figs.~\ref{CdSnTe-band}(c) and 
\ref{CdSnTe-band}(d)~\cite{rashba}. 
The RSS  is  enhanced in d-CdSnTe due to large Z value of Te,  as well as due to lack of mirror symmetry in the nanoribbon. 
 As the width of d-CdSnTe is increased from 2$m$ to 5$m$,  the band separation at the $\Gamma$ point,  near to Fermi energy, is observed to decrease and the bands tend to overlap.  To confirm that the band splitting in d-CdSnTe nanoribbons is of Rashba type, we plotted spin resolved band structure of d-CdSnTe nanoribbon for $5m$ width after inclusion of SOC and shown in Fig. \ref{CdSnTe-spin-split}. Due to inclusion of SOC, the spin up and spin down bands show momentum dependant splitting indicating the band splitting is of Rashba type. At the $\Gamma$ symmetry line, these bands majorly show $S_{y}$ and $S_{z}$ character as shown in Fig. \ref{CdSnTe-spin-split}(b) and \ref{CdSnTe-spin-split}(c).
  Our analysis shows that  RSS of the conduction band at $\Gamma$ point is unaltered for all 
the four nanoribbon widths, we have studied.
Since the Rashba splitted bands occur very close to the Fermi energy,  Zeeman-type as well as  Rashba   splitting is seen in d-CdSnTe nanoribbons. With  modifications in the Fermi energy at the Rashba 
split band by applying external voltage, 
non-dissipative spin current can be obtained and thus these nanoribbons can have applications in spintronics devices. 
Spin density plots in Figs.~\ref{CdSnTe-Spin}(c) and \ref{CdSnTe-Spin}(d) show that the edge terminating with Sn atom
 and its adjacent Te atoms is contributing to the magnetic moment of the cell almost similar to the situation without inclusion of SOC.

 The Rashba splitting is characterized by the Rashba energy $E_R$ in eV 
 and the corresponding momentum offset $\Delta k$  away 
 from the crossing point as elaborated in Fig.~\ref{spin}.  The RSS
 coefficient is defined as $\alpha_R = 2E_R/\Delta k$.  It can be 
 inferred that  $\alpha_R$ 
 arises due to Te atoms only.
Table~\ref{rashba} lists all the Rashba parameters and our calculated values 
are of same order of magnitudes as reported in the previous works~\cite{tao,Mao} for Janus transition metal dichalcogenide monolayers and Pb-adsorbed  WSe$_2$ monolayer. Such simultaneous occurrence of Zeeman-type
 and Rasbha spin splitting is due to the lack
 of inversion symmetry which is  observed previously in Sn triangular lattice
 atomic layer~\cite{yaji} and Pb-adsorbed WSe$_2$ monolayer~\cite{Mao}.  
 In d-CdSnTe nanoribbons, both RSS and Zeeman-type 
 spin splitting occur near the Fermi level  around $\Gamma$ point.  Large Z value and the  local
 orbital angular momentum
of the edge atoms  play a crucial role in RSS
splitting along with asymmetric charge distribution on inclusion of SOC~\cite{park}.

   To summarise the properties of distorted armchair CdSnX (X = S, Se and Te) nanoribbons,  we observe that  the systems exhibit nonzero magnetic moments localized on Sn and X atoms due 
 to their edge terminating positions that reduces the co-ordination of respective atoms. The spin density giving rise to localized magnetic moments changes with the width of the nanoribbon and the chalcogenide atoms present in the nanoribbons as a result of variations in the interaction and bonding among the atoms. 
 Our calculated values of the magnetic moments of the unit cells for the distorted edge nanoribbons (The Table S3 in the Supplementary information) and are in good agreement with the magnetic moments for metallic transition
metal dichalcogenides reported by 
 Li~\textit{et al.}~\cite{li3}.  
 For d-CdSnSe nanoribbons,  the other edge 
 with Se atoms bonded to Cd atoms is not showing any spin 
 density for larger widths after inclusion of SOC while for d-CdSnTe nanoribbons,  the other edge 
 with Te atoms bonded to Cd atoms does not show any spin density for all widths with and without inclusion of SOC.
 
 The opening of band gap at two time irreversible $\vec{k}$-points on or near Fermi level for S, Se and Te case mainly  occurs due to the inclusion of SOC on Sn atoms while the Zeeman-type or Rashaba spin splitting at the $\Gamma$ point is mainly due to inclusion of SOC in the calculations for chalcogen atoms Se and Te. Figure~\ref{spin} nicely depicts the change in the spin splitting as we change the X atoms. Only band inversion is seen with no spin splitting for  X = S nanoribbons with presence of two Weyl points near the Fermi energy. Zeeman-type spin splitting along with band inversion is observed in X = Se nanoribbons along with opening of slightly larger gaps than X = S nanoribbons on inclusion of SOC in the calculations. Rashba spin splitting along with Zeeman-type splitting is observed in X = Te nanoribbons. Band inversion and gap openings on inclusion of SOC are also clearly depicted in the figure.

\subsection{Normal armchair edge CdSnX Nanoribbons}
As shown in Fig.~\ref{mon}(e),  one edge of the armchair nanoribbons has Cd and X atoms while the other edge is formed by Sn and X atoms.  Unlike distorted CdSnX nanoribbons,  all the normal armchair nanoribbon structures show 
zero magnetic moment.  This indicates that the exchange field is absent in these
systems.  The Sn $p$ and X $p$ states split into $p_{x}$, $p_{y}$ and $p_{z}$ constituent states 
indicating the presence of crystal field splitting in the system.

We have systematically increased the width of the nanoribbons
from 2$m$ to 
5$m$  for all the X atoms of the chalcogen in a-CdSnX (X = S, Se and Te) nanoribbons.  All the nanoribbons show
semiconducting property for all widths. The band gaps for all a-CdSnX (X = S, Se and Te) systems are listed in Table~\ref{gap}.
 
\begin{figure}
{\includegraphics[scale=0.43, valign=t]{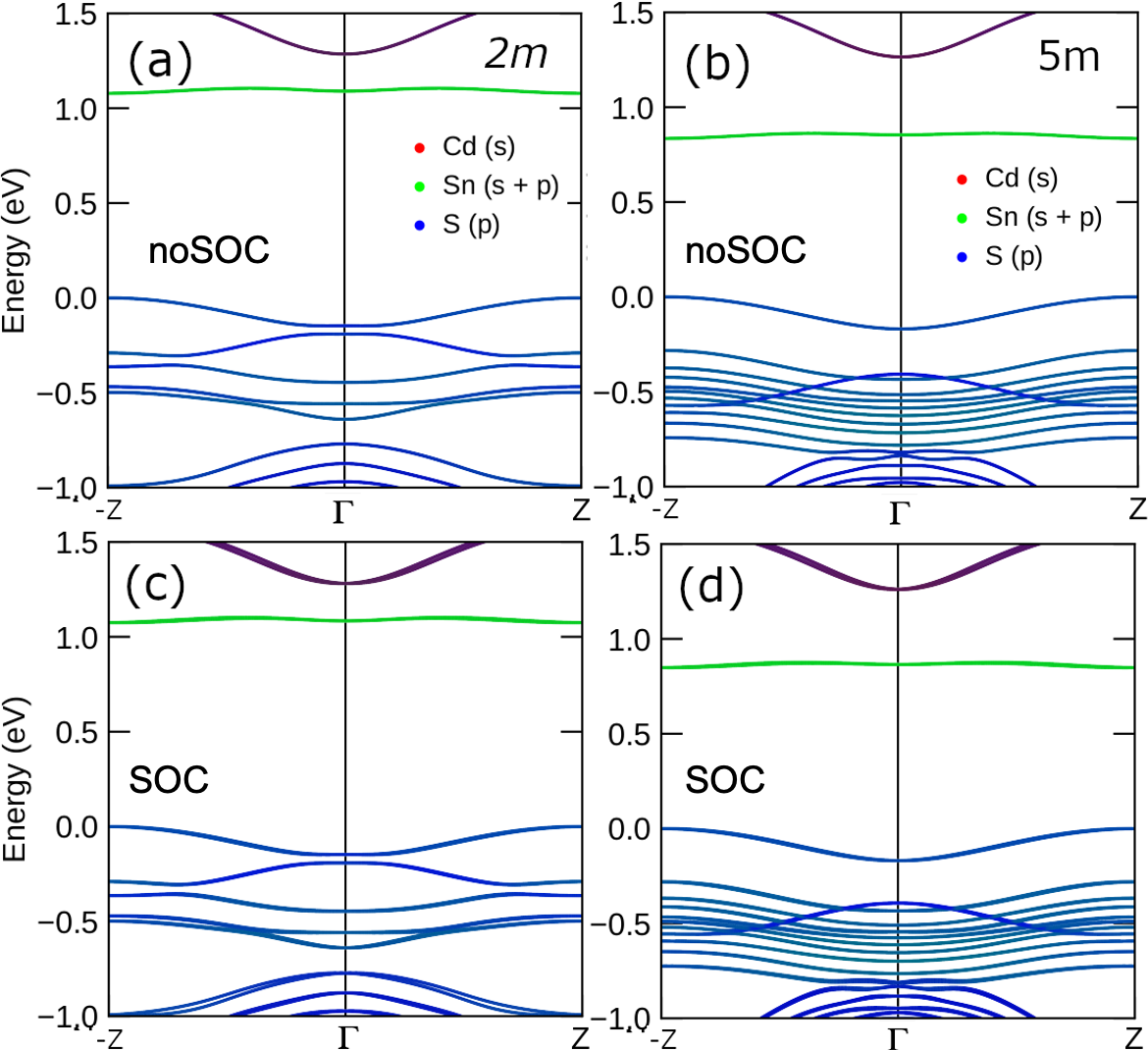}}   
   \caption{Band structure plots for a-CdSnS nanoribbons without SOC and with SOC for widths 2$m$ and 5$m$ are shown in (a) and (c) and  (b) and (d) respectively.  }
  \label{CdSnS-armchair}
  \end{figure}

  \begin{figure}
{\includegraphics[scale=0.43, valign=t]{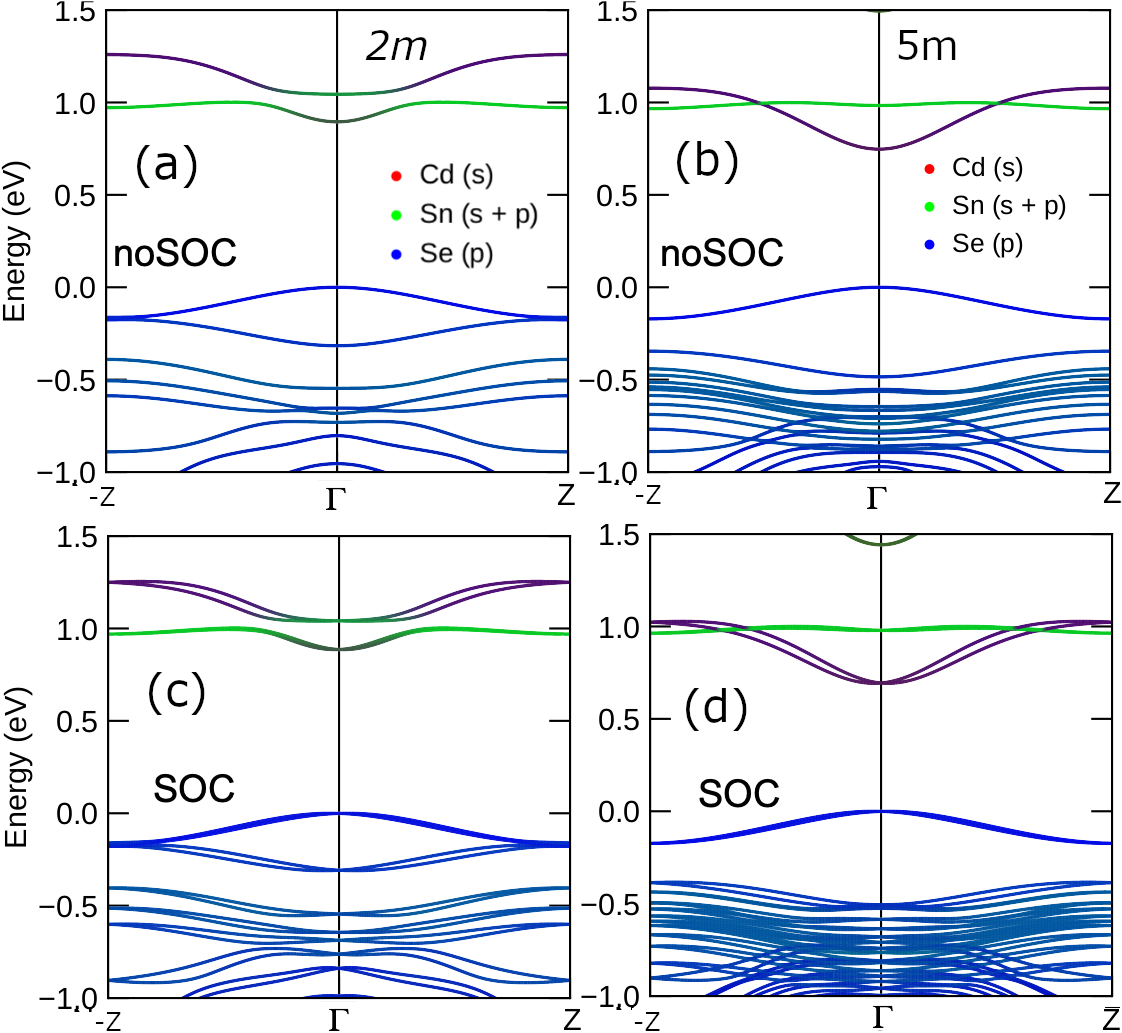}}    
   \caption{Band structure plots for a-CdSnSe nanoribbons without SOC and with SOC for widths 2$m$ and 5$m$ are shown in (a) and (c) and  (b) and (d) respectively.  }
  \label{CdSnSe-armchair}
  \end{figure} 

 As the width is varied from 2$m$ to 5$m$ for a-CdSnS nanoribbons (Figs.~\ref{CdSnS-armchair}(a) and \ref{CdSnS-armchair}(b)), the direct band gap changes to indirect band gap. The bands near Fermi energy arise due to orbital
hybridization between Sn $p$ and S $p$ states. 
After inclusion of SOC,  the  bands show weak splitting away from the $\Gamma$ point, specifically  at Z and -Z symmetry points, more so for nanoribbons of width 5$m$.  This forms valleys as revealed from Figs.~\ref{CdSnS-armchair}(c) and \ref{CdSnS-armchair}(d). 
  \begin{figure}
  {\includegraphics[scale=0.46, valign=t]{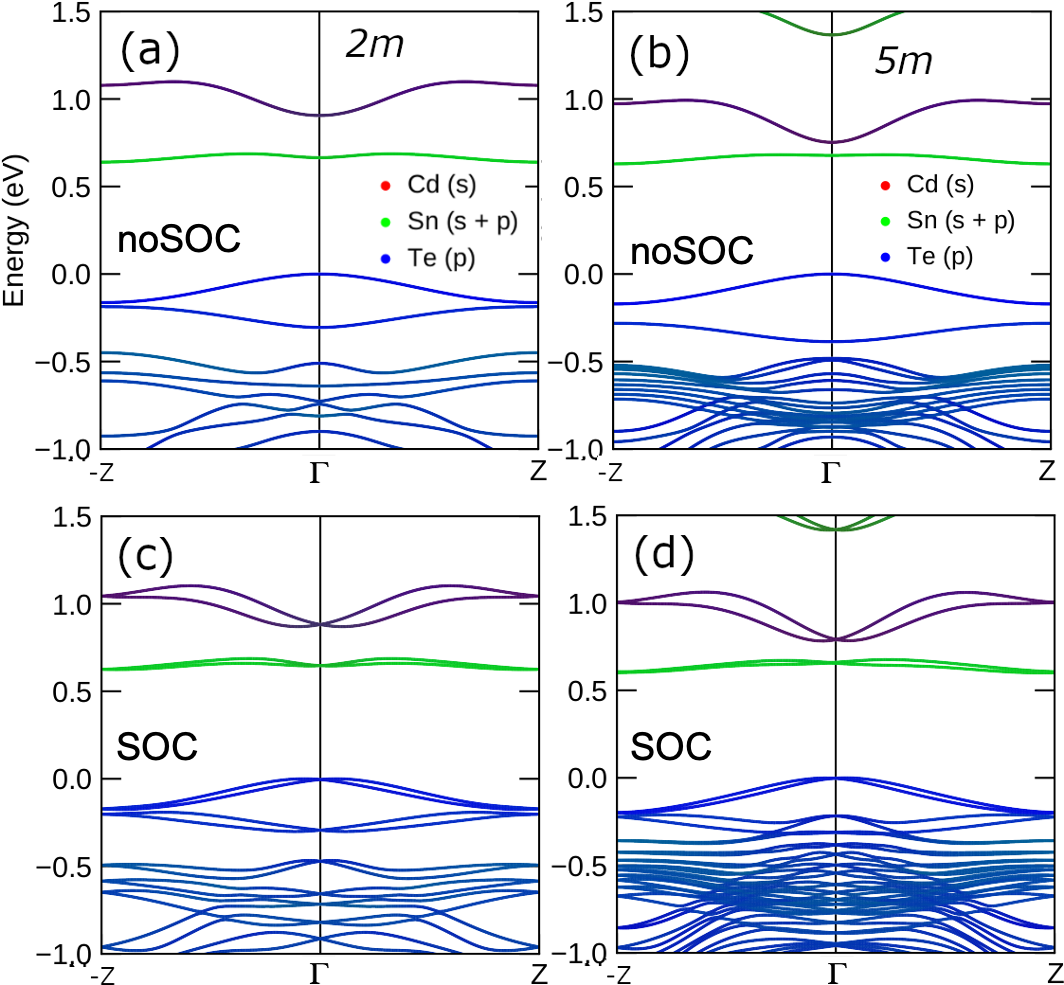}}   
   \caption{Band structure plots for CdSnTe armchair nanoribbons for widths 2$m$ and 5$m$ are shown in  (a) and (b)~without inclusion of SOC and in (c) and (d)~with SOC respectively. }
  \label{CdSnTe-armchair}
  \end{figure}

As X is changed from S to Se in the a-CdSnX nanoribbons, we  observe that the band inversion takes place 
in the valence and conduction bands as shown in Figs.~\ref{CdSnSe-armchair}(a) and \ref{CdSnSe-armchair}(b).    Valence 
band region also imprints very strong hybridization between Sn $sp$ and Se $p$ states, while in the conduction band region Cd $s$ and
Se $p$ orbital hybridization is observed.Application of SOC reveals that the  magnitude of the spin splitting  is more in a-CdSnSe (Figs.~\ref{CdSnSe-armchair}(c) and \ref{CdSnSe-armchair}(d))
than that of a-CdSnS  nanoribbons (Figs.~\ref{CdSnS-armchair}(c) and \ref{CdSnS-armchair}(d)). The band crossings are mainly due to Se $p_x$ and $p_z$ states and Sn $p_x$ and  $p_z$ states. Band structure plots show that a-CdSnSe nanoribbons exhibit Rashba spin splitting in the conduction bands (clearly visible for nanoribbon of width 5$m$) which was not observed in d-CdSnSe nanoribbons.

\begin{table}
 \setlength{\tabcolsep}{15pt}
\renewcommand{\arraystretch}{1.6}
 \caption{Band gap values for CdSnX (X = S, Se, Te) armchair edge nanoribbons for widths 2$m$ to 5$m$.}
 
 \begin{tabular}{c c c c }
 \hline\hline
    Width & CdSnS & CdSnSe &CdSnTe\\
  \hline\hline
  2$m$ &1.078 & 0.895&0.639\\
  
  3$m$ &1.028 &0.858 & 0.631\\
  
  4$m$ & 0.951 &0.803 &0.629\\
  
  5$m$ & 0.835 &0.747&  0.629\\ 
  \hline
 \end{tabular}
\label{gap}
\end{table}


  \begin{figure}
{\includegraphics[scale=0.35, angle=270]{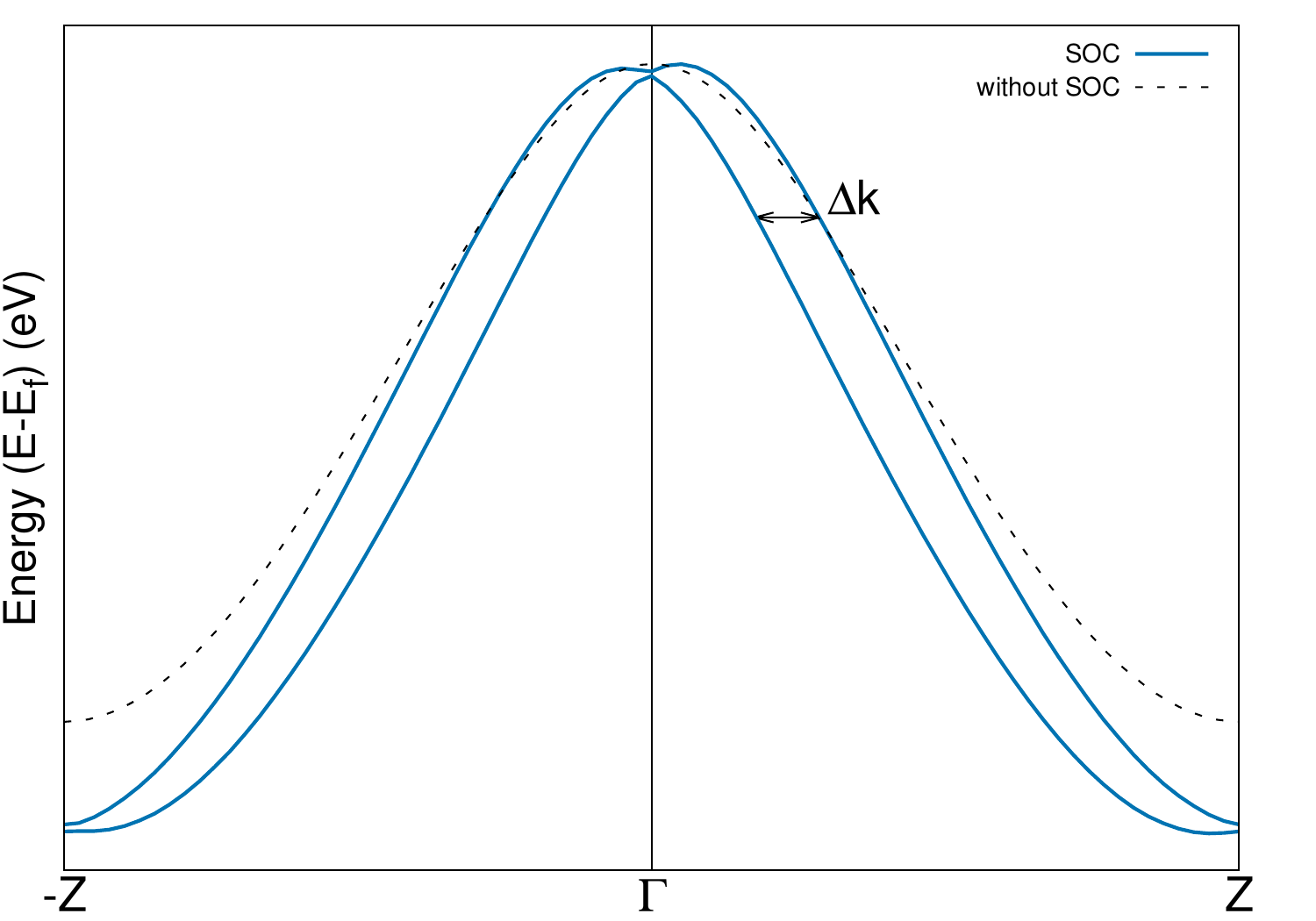}}   
   \caption{The existence of Rashba splitting in CdSnTe armchair nanoribbon for 5$m$ width. The band splitting  due to SOC (continuous line) along $\mathbf{k}$ is compared with degenerate bands without inclusion of  SOC (dotted line).}
  \label{rashba-band}
  \end{figure}

a-CdSnTe system turns into indirect band gap 
semiconductor as shown in Figs.~\ref{CdSnTe-armchair}(a) and \ref{CdSnTe-armchair}(b) with a relatively smaller band gap value.
Strong hybridization of Sn $s$ and $p$ and Te $p$ states is seen in these
systems.  The inclusion of SOC splits the bands into spin 
components even more than a-CdSnSe nanoribbons owing to high Z value of Te.  The Rashba spin splitting is
visible in a-CdSnTe nanoribbons in the conduction band at the $\Gamma$ point in Figs.~\ref{CdSnTe-armchair}(c) and \ref{CdSnTe-armchair}(d). Figure~\ref{rashba-band} gives the 
enlarged view of Rashba splitting in armchair nanoribbons.

\subsection{Zigzag Nanoribbons}
\begin{figure}
{\includegraphics[scale=0.44]{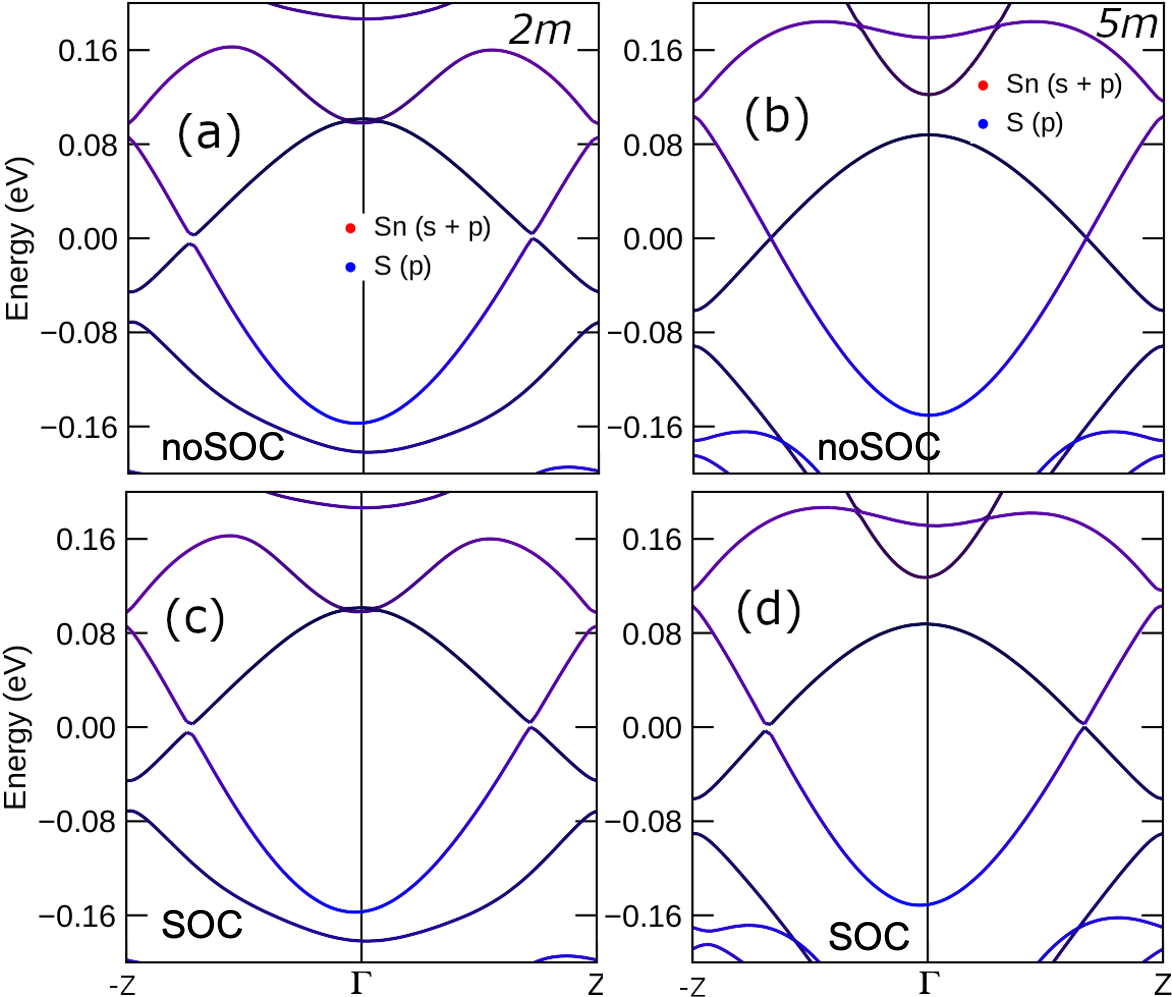}} 
   \caption{Band structure plots for z-CdSnS nanoribbons 
   of widths 2$m$ and 5$m$ without and with inclusion of SOC are shown in (a) and (c) and (b) and (d) respectively.}
  \label{CdSnS-zigzag}
  \end{figure}

\begin{figure}
{\includegraphics[scale=0.42]{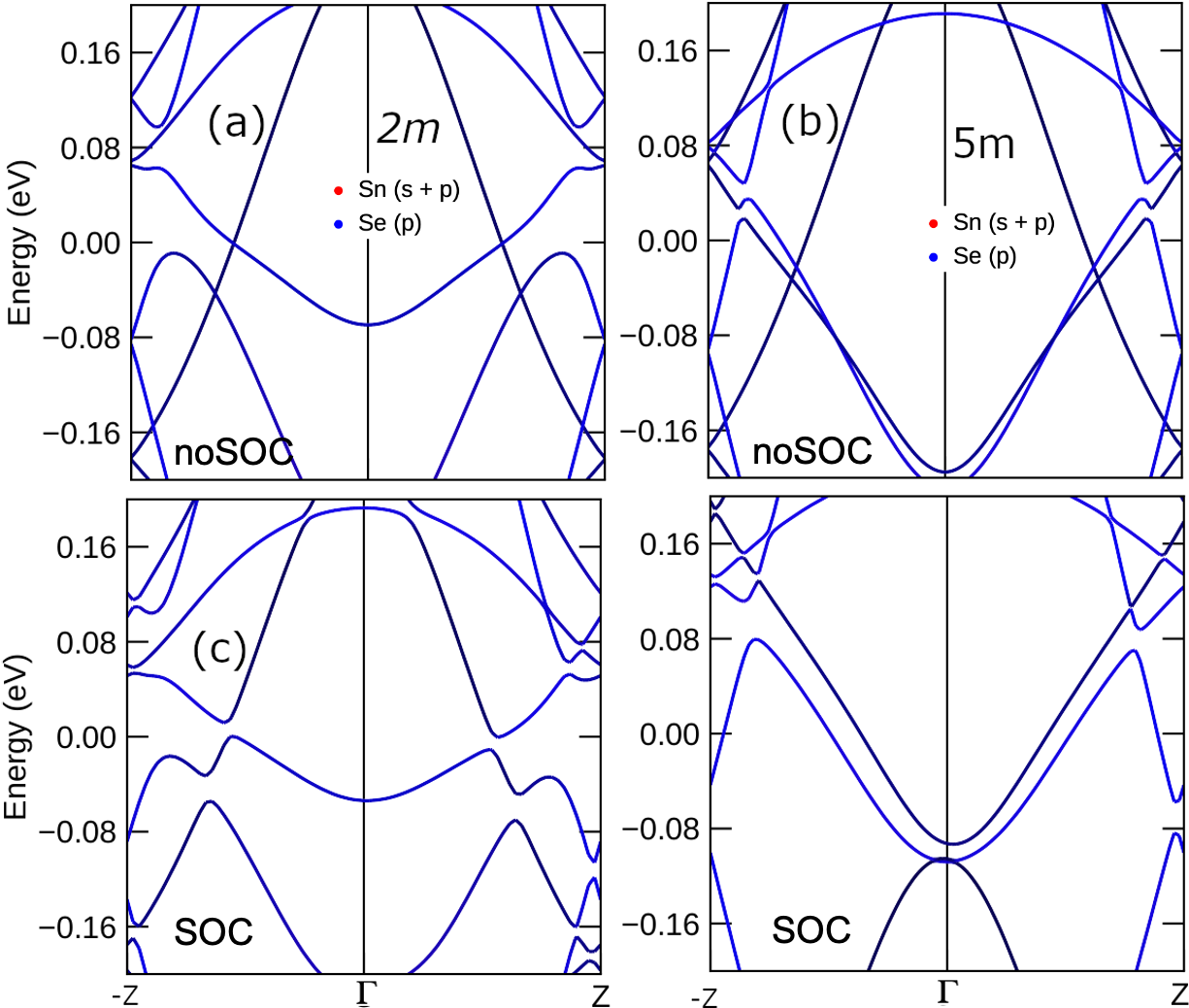}} 
   \caption{Band structure plots for z-CdSnSe nanoribbons 
   of widths 2$m$ and 5$m$ without and with inclusion of SOC are shown in (a) and (c) and (b) and (d) respectively.}
  \label{CdSnSe-zigzag}
  \end{figure}

\begin{figure}
{\includegraphics[scale=0.42]{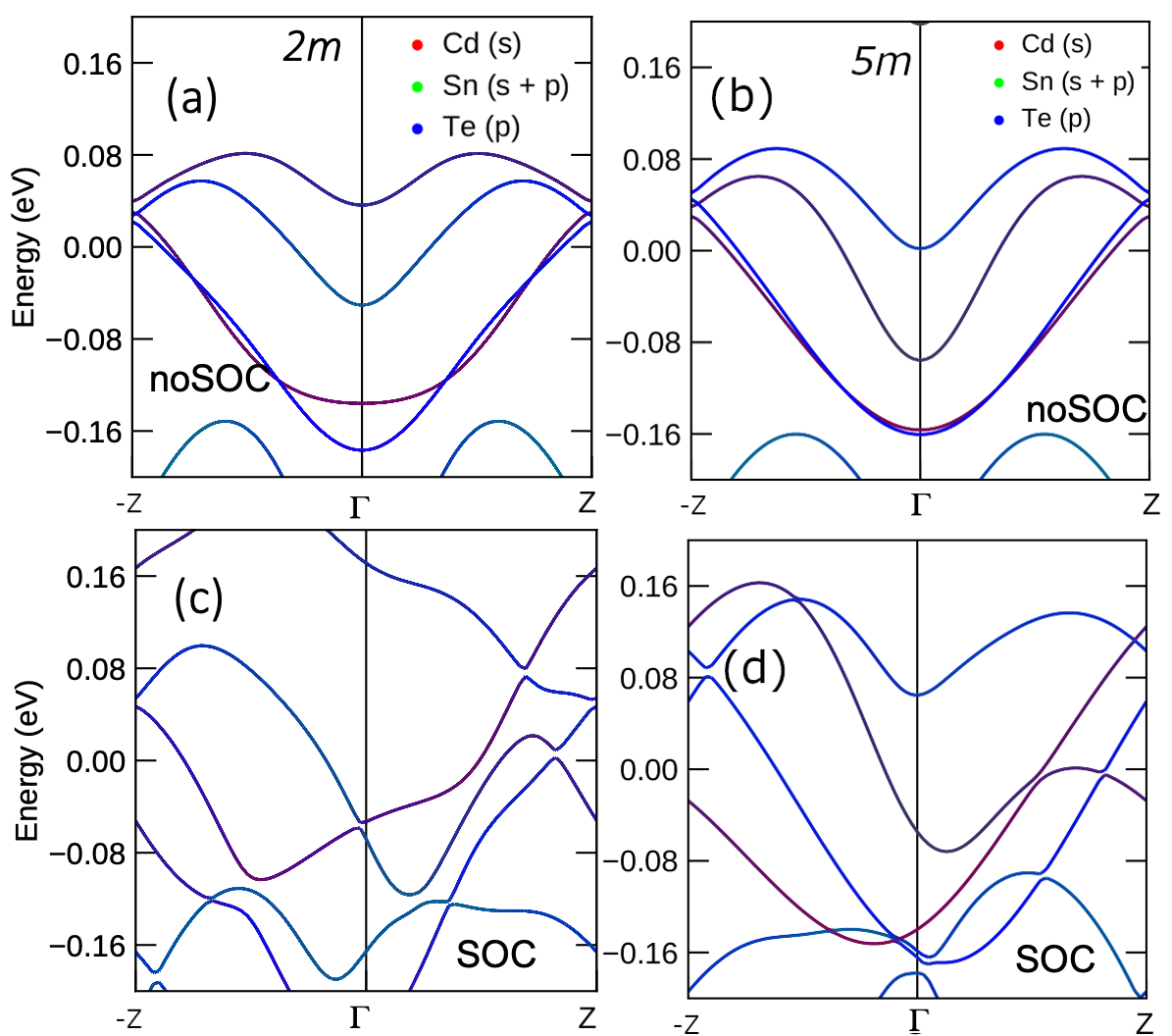}} 
   \caption{Band structure plots for z-CdSnTe nanoribbons 
   of widths 2$m$ and 5$m$ without and with inclusion of SOC are shown in (a) and (c) and (b) and (d) respectively.}
  \label{CdSnTe-zigzag}
  \end{figure} 

For completeness, we have performed the electronic structure calculations for z-CdSnX (X = S, Se and Te) nanoribbons shown in Fig.~\ref{mon}(d). The geometric structure of these nanoribbons has inherent mirror symmetry along the nanoribbon length, although  the two edges are not identical, and hence the symmetry effects are reflected in their band structures. 

The electronic band structures of z-CdSnS nanoribbons for widths 2$m$ and 5$m$ are shown in Figs.~\ref{CdSnS-zigzag}(a) to \ref{CdSnS-zigzag}(d).  The band structures exhibit time reversal symmetry with small band gap opening at the Fermi energy even without inclusion of SOC.  After inclusion of SOC, the band structure remains mostly unaffected, except increased gap at Fermi energy, as the effect of SOC is due to Sn atoms only.  As the width is increased, the edge atom interaction is reduced which reduces the band gaps.  The band structures for all widths show signs of band inversion character. The bands at the Fermi energy are mainly due to S $p$ states. 

The z-CdSnSe nanoribbons are metallic irrespective of the width of the nanoribbon for calculations without inclusion of SOC.  However, the effect of SOC is seen in larger z-CdSnSe nanoribbons in comparison to that of z-CdSnS nanoribbons. The band gap openings are prominent at time reversal symmetry points. After inclusion of SOC, the band gap at the band crossing for nanoribbon of width 2$m$ opens up at Fermi energy with the signature of band inversion as shown in Fig.~\ref{CdSnSe-zigzag}(c). The z-CdSnSe nanoribbon of width 5$m$ shows semi-metallic character. 
In comparison to the z-CdSnS nanoribbons, z-CdSnSe nanoribbons show less band inversion character.

  \begin{figure}
{\includegraphics[scale=0.35, angle=270]{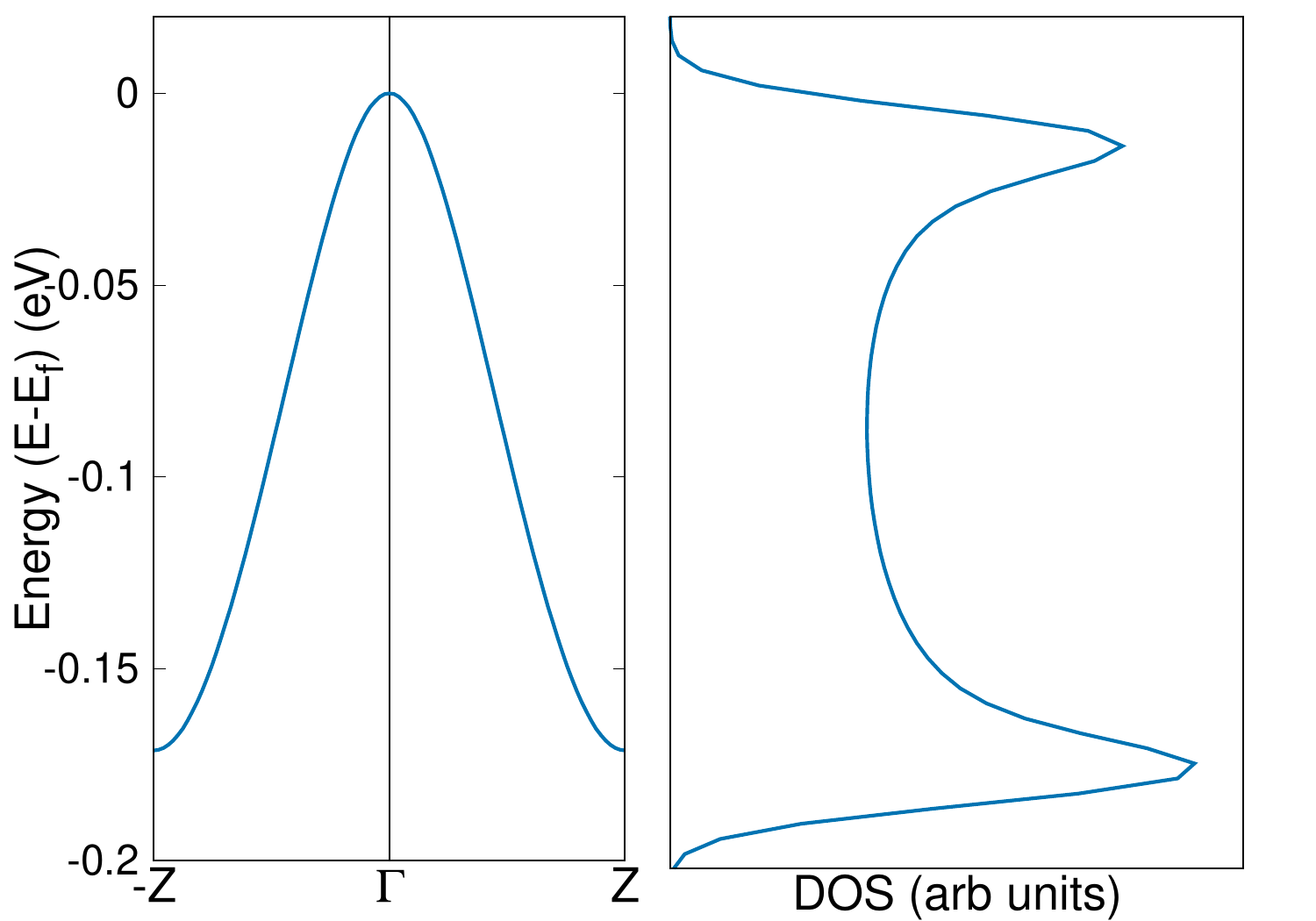}}   
   \caption{Van-Hove singularities in DOS are shown for armchair nanoribbon of width 2$m$ in the right panel.  The equivalent saddle points in the band structure are shown in the left panel. }
  \label{van-hov}
  \end{figure}

The electronic band structures of z-CdSnTe nanoribbons exhibit metallic nature as can be seen from Fig.~\ref{CdSnTe-zigzag} for all widths irrespective of inclusion of SOC. The bands near the Fermi energy mainly arise from Te $p$ states. The band structures for z-CdSnTe nanoribbons exhibit absence of time reversal symmetry along Z-$\Gamma$-(-Z) lines.

All the zigzag edge nanoribbons have cell magnetic moment more than the corresponding distorted edge nanoribbons. The magnetic moment in zigzag nanoribbons  arises due to the edge Sn and X atoms.

\section{Conclusion}\label{Discussions}

The  electronic structure study of nanoribbons  prepared 
 from Sn doped CdX (X = S, Se, Te ) monolayers 
 is carried out in this piece of work.
 We considered mainly three types of nanoribbons viz.,(a)~distorted
 armchair edges, (b)~normal armchair  edges and (c)~ normal zigzag edges.

 Lack of inversion and mirror symmetry in these systems, along with the presence of strong 
 exchange field, induces
  spin splitting in absence of external
 magnetic field.  The valence and conduction bands 
 cross the Fermi energy
 at two conjugate time reversible points for all
 the d-CdSnX nanoribbons.  
 Inclusion of SOC opens a band gap (not at Fermi energy) upto few meV at the band crossings
 with the formation of electron and hole
 pockets keeping the metallic
 characters intact for all
 the nanoribbons.  The semiconducting monolayers thus
 turn into metallic and magnetic nanoribbons,  due 
 to the size confinement and nature of the edge atoms.  
 Such types
 of changes in the band structures are also seen previously in 
 MoS$_2$ nanostructures and InSe nanoribbons~\cite{DAVEL,Yao}.
 
 For  the d-CdSnS nanoribbons,  we observe Weyl semi-metal type band structure irrespective of the width of the nanoribbons.  
 The band gap opens up in the Weyl cone due to splitting in addition to 
 tilting of the bands after inclusion of SOC.
  The d-CdSnSe nanoribbons are more interesting as they show spin splitting
 due to the exchange field
  for smaller width nanoribbons
 and Zeeman-type spin splitting for
 widths 4$m$ and 5$m$.  The d-CdSnTe nanoribbons show   
  Rashba spin splitting  around $\Gamma$ point with band tilting due to high Z value for Te.


 The different types of spin splittings are consistent with the presence of bulk inversion asymmetry and structure inversion 
 asymmetry~\cite{Winkler}.  The splitting is
 also dependent on the local orbital angular moment which 
 changes as we change the X atoms from S to Te due to the effect of SOC~\cite{park}.

One very striking feature of all the d-CdSnX nanoribbons,  is that the bands 
crossing the Fermi level open up  revealing a SOC gap at  time irreversible
$\vec{k}$-points near the Fermi energy. This opening of
the gap is purely contributed by SOC of Sn atom  while the band splittings around the
$\Gamma$ point is due to the SOC of the chalcogen atoms.
As the SOC of Se and Te is more than S, those systems show Zeeman-type and Rashba
spin splitting around the $\Gamma$ point.  For distorted armchair edge nanoribbons the spin splitting due to SOC is
observed more in unoccupied states in valence band 
region while for normal armchair case the same is observed
more in conduction band region. Distorted armchair nanoribbons have non-zero cell magnetic moments while normal armchair nanoribbons have zero magnetic moment.

Spike-like discontinuous states in the DOS, referred to as 
van Hove singularities
(vHs), are visible for all the distorted and armchair edge nanoribbons 
revealing confinement of electrons and phonons along one dimension~\cite{zhang1}. In Fig. \ref{van-hov}, the van-Hove singularities for armchair nanoribbon for $2m$ width is shown as an example. These 
spikes in DOS have very important role in optical transitions.

All the armchair CdSnX nanoribbons remain semiconducting 
with mainly Zeeman-type spin splitting and Rashba spin splitting for
X = Se and Te systems respectively. In each case, the unit cell acquires no magnetic moment.  A recent study has shown that without external
 magnetic field Zeeman-type term ($H_Z$) (called as exchange term) 
 can induce spin splitting~\cite{Qiao}. The   armchair CdSnTe nanoribbons
can be potential candidates for superconductivity if the 
Fermi level can be properly optimized around the Rashba spin splitted band~\cite{floq,loder}.

This work has established that CdSnX nanoribbons 
do possess interesting and exotic electronic
structure. Confirmation of the topological
significance has to be established in terms of presence of edge states with their chirality,
finite Chern numbers, anomolous hall current, spin texture etc. We plan to extend the present work for such studies.

\section{Acknowledgements}
SC acknowledges financial support from
Department of Science and Technology, Government of India through the
Women Scientist-A (WOS-A) program.  PARAM-BRAHMA super computer under 
National Supercomputing Mission established at IISER,  Pune  has been used for 
the DFT calculations and the support is gratefully acknowledged.

\bibliography{refs1}
\end{document}


\title{Effect of Exchange Interaction and Spin-Orbit Coupling on Spin Splitting in CdSnX (X = S, Se ad Te) nanoribbons}

\author{Sutapa Chattopadhyay}
 \affiliation{Department of Physics, Savitribai Phule Pune University, Pune 411 007, India}
\author{Vikas Kashid}%
 \affiliation{Department of Physics, Savitribai Phule Pune University, Pune 411 007, India}%
 \affiliation{MIE-SPPU Institute of Higher Education, Doha -- Qatar}
\author{P. Durganandini}
 \affiliation{Department of Physics, Savitribai Phule Pune University, Pune 411 007, India}
\author{Anjali Kshirsagar}
\affiliation{Department of Physics, Savitribai Phule Pune University, Pune 411 007, India}
\maketitle

\begin{center}
\textbf{\Large Supplementary Information }
\end{center} 

{\bf Table S1.} Lattice parameters ($a = b$) of pristine CdX monolayers along with the 
buckling index $d_z$ and Cd-X
bond lengths.
\vskip -0.2in
\begin{table}[h!]
\setlength{\tabcolsep}{15pt}
\renewcommand{\arraystretch}{1.8}
\centering{
\vspace{0.1in}
\begin{tabular}{ | c | c | c | c | }
\hline
Properties $\downarrow$&  \multicolumn{3}{c|}{Pristine CdX monolayers} \\
 \hline
  & CdS & CdSe & CdTe\\
\hline
$a(= b)$ & 4.11  & 4.43  & 4.72  \\
 \hline
$d_z$  & 0.0   & 0.0   & 0.0  \\
 \hline
Cd-X &  2.52  & 2.58 & 2.77  \\
\hline
\end{tabular}}
 \end{table}

\vskip 0.2in

{\bf Table S2.} Lattice parameters ($a = b$) of Sn-doped CdX monolayers along with the 
buckling index $d_z$ and Cd-X and Sn-X bond lengths .

\vspace{-0.05in}
\begin{table}[h!]
\setlength{\tabcolsep}{15pt}
\renewcommand{\arraystretch}{1.8}
\centering{
\begin{tabular}{ | c | c | c | c | }
\hline
 Properties $\downarrow$ &   \multicolumn{3}{c|}{Sn-doped CdX monolayers}\\
    \hline
  & CdS & CdSe & CdTe\\
\hline
$a(= b)$ (in \AA ) &  3.99 & 4.16 & 4.45 \\
 \hline
$d_z$ (in \AA ) & 1.25 & 1.26  & 1.29 \\
\hline
Cd-X (in \AA ) &  2.47 & 2.78 & 2.78\\
\hline
Sn-X (in \AA ) & 2.58, 2.64 & 2.72, 2.78 & 2.90, 2.98 \\
\hline
\end{tabular}}
\end{table}

\newpage

{\textbf{\large{Modeling and Geometric Structure of Nanoribbons}}}

\vspace{0.2in}

We have modeled CdSnX nanoribbons by cutting Sn doped CdX monolayers parallel to lattice vector
⃗$\vec{a}$ as shown in Fig. 1(b) in the main text of paper and parallel to lattice vector $\vec{b}$ as shown in Fig. 1(c). For
periodic calculations, vacuum is introduced along $\vec{b}$ and $\vec{a}$ respectively in these two structures. The open
edges of nanoribbons form distorted armchair structure as shown in Fig. 1(b) (Top edge has Cd atoms in
the first row, X atoms in the second row and all three atoms in the armchair while the bottom edge has
Sn atoms in the first row and all three atoms in the double armchair, the width of the nanoribbon being in
a direction $\perp$ to $\vec{a}$ $i.e.$, $\perp$ to the infinite direction. Conversely, if the monolayer is sliced parallel to $\vec{b}$ and
vacuum is introduced along $\vec{a}$, the resultant nanoribbon is infinite along $\vec{b}$, with the open edges forming
distorted armchair structure at the left edge and right edge as shown in Figs. 1(c). Thus the nanoribbons
infinite along $\vec{b}$ (Fig. 1(c)) have the same structure as nanoribbons infinite along $\vec{a}$ (Fig. 1(b)) with their
left edge being the mirror image of the top edge and right edge being the mirror image of the bottom
edge. The two nanoribbons are therefore expected to show same physical and electronic properties.

\vspace{0.2in}

If the Cartesian x-axis and lattice vector $\vec{a}$ are aligned for the distorted edge nanoribbon, then y-axis
is at 30$\textdegree$
from the lattice vector $\vec{b}$ making these nanoribbons tilted.

\vspace{0.2in}

We have also generated nanoribbons with normal armchair and normal zigzag edges along finite sides
by reorienting the monolayers that generate nanoribbons which are infinite along Cartesian axes (unlike
tilted nanoribbons). These zigzag and armchair nanoribbons also have different atoms at their two edges
as shown in Figs. 1(d) and 1(e) respectively.

\vspace{0.2in}

The vacuum separation between two nanoribbon images is kept $\geq$ 15~\AA\  in the two periodic directions  ̊perpendicular to their lengths to avoid the interactions between the periodic images.

  \begin{table}[h!]
 \setlength{\tabcolsep}{15pt}
\renewcommand{\arraystretch}{1.6}
 {\bf Table S3.} Magnetic moments of the unit cell for all the widths of d-CdSnX (X = S, Se, Te) nanoribbons in $\mu_{B}$. \\
 \centering{}
 \begin{tabular}{c c c c}
  \hline \hline
     Width&CdSnS&CdSnSe&CdSnTe\\
   \hline\hline
   2$m$ &1.19&1.17&0.62\\
   
   3$m$ &1.27&1.26&0.83\\
   
   4$m$ & 1.30 &1.29& 0.66\\
   
   5$m$ &1.31&1.09&0.68\\
   \hline
 \end{tabular}
 \label{mag}  
 \end{table}




\newpage 
\begin{figure}
\hspace{0.9in}
{\includegraphics[scale=0.5]{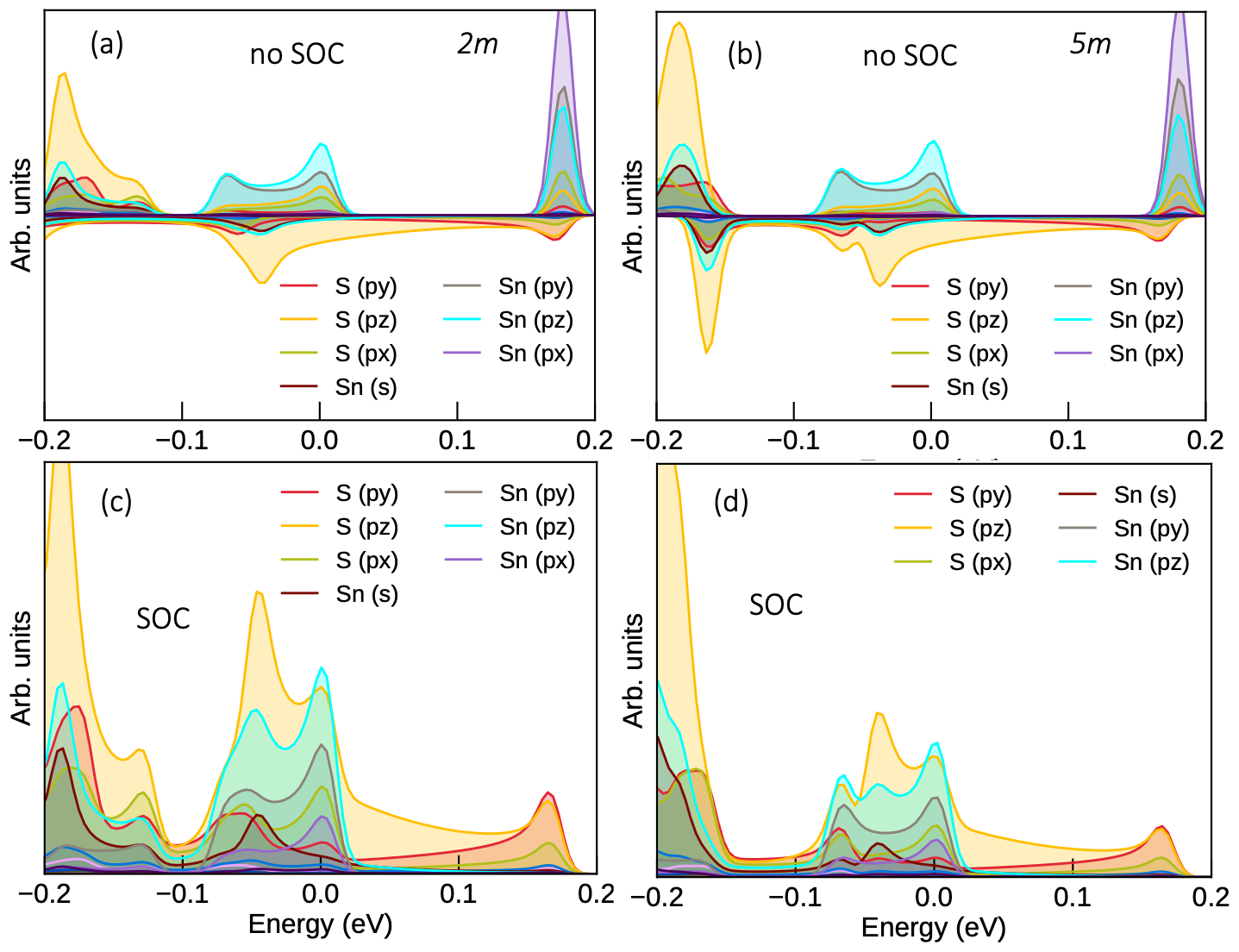}}
\vskip -0.1in
 {\bf Figure S1.} The partial density of states (PDOS) plots for d-CdSnS nanoribbons with respective Fermi energy set to zero.  The left and right panels indicate the PDOS of the nanoribbons with widths 
 2$m$ and 5$m$ respectively with the upper panels for calculations without inclusion of SOC and the bottom panels with  inclusion of SOC. Upper panels show spin polarized PDOS.
\end{figure}

 \begin{figure}
{\includegraphics[scale=1.05]{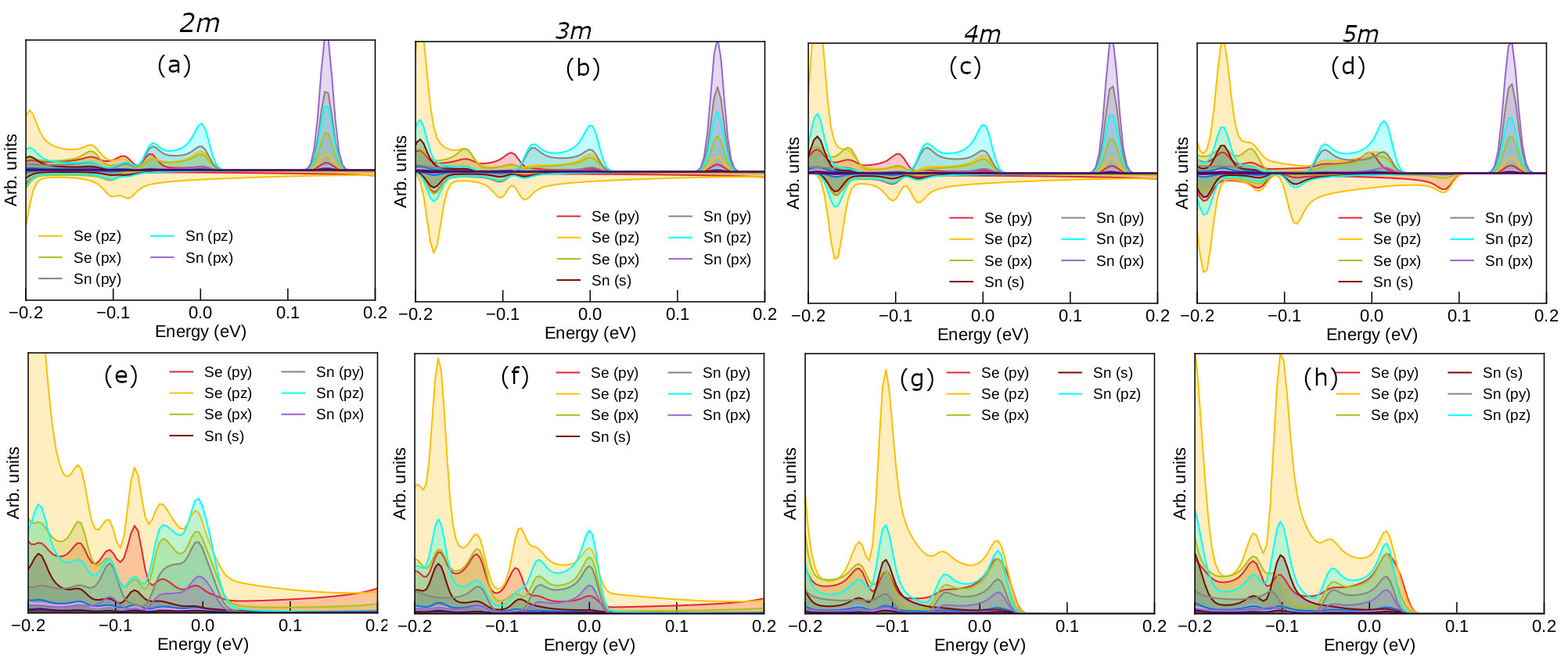}}  
{\bf Figure S2.} PDOS plots for d-CdSnSe nanoribbons with widths ranging from 2$m$ to 5$m$ with the respective Fermi energy set to zero.  Panels (a) to (d) indicate the PDOS without inclusion of SOC in the calculations while panels (e) to (h) show PDOS after inclusion of SOC.  Panels (a)-(d) show spin polarized PDOS.
\end{figure}

\begin{figure}
\begin{center}
 {\includegraphics[scale=0.45]{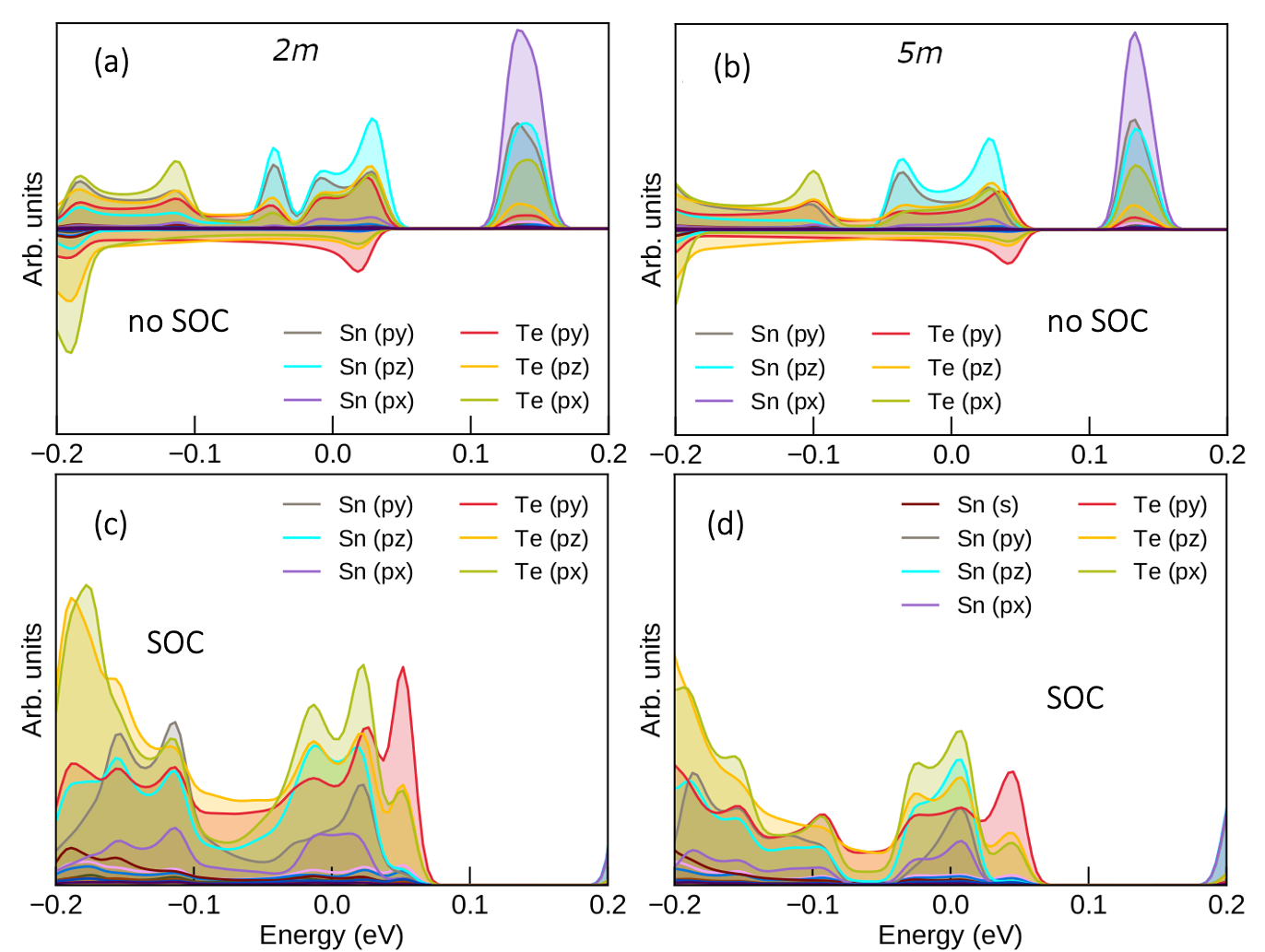}} 
 \end{center}
 \vskip -0.1in
{\bf Figure S3.} PDOS plots for d-CdSnTe nanoribbons for widths 2$m$ and 5$m$ with the respective Fermi energy set to zero. Top panels are for calculations without inclusion of SOC and the bottom panels are  with inclusion of SOC. Upper panels show spin polarized PDOS.
\label{CdSnTe-PDOS}
\end{figure}
\begin{figure}
\begin{center}
{\includegraphics[scale=0.5, valign=t]{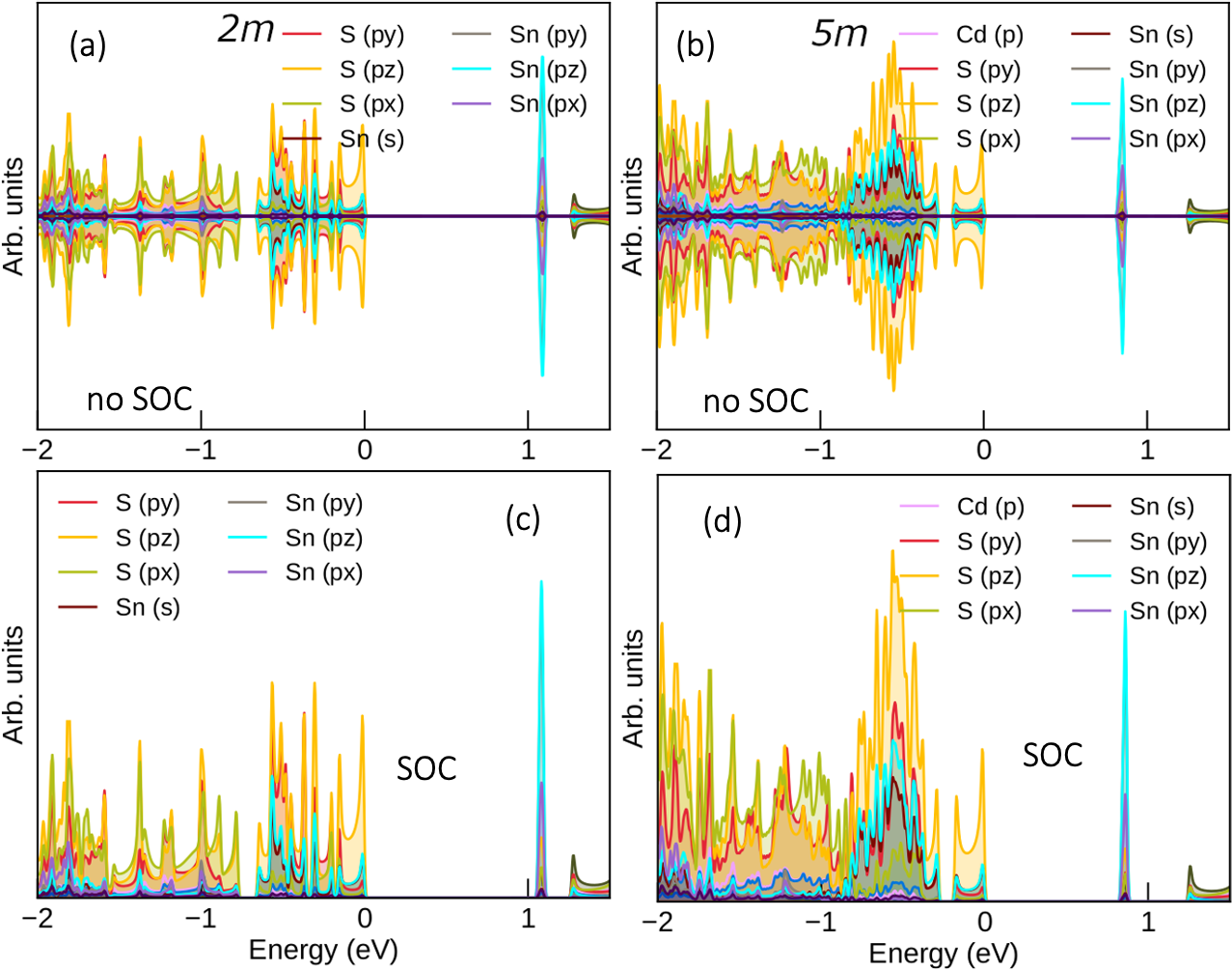}}
\end{center}
{\bf Figure S4.} PDOS plots for CdSnS armchair edge nanoribbons, with the respective Fermi energy set to zero.
Plots for width 2$m$ without SOC and with SOC are shown in the left top and left bottom panels whereas for width 5$m$ are shown in right top and right bottom panels. Upper panels show spin polarized PDOS.
\end{figure}

\begin{figure}
\begin{center}
{\includegraphics[scale=0.45]{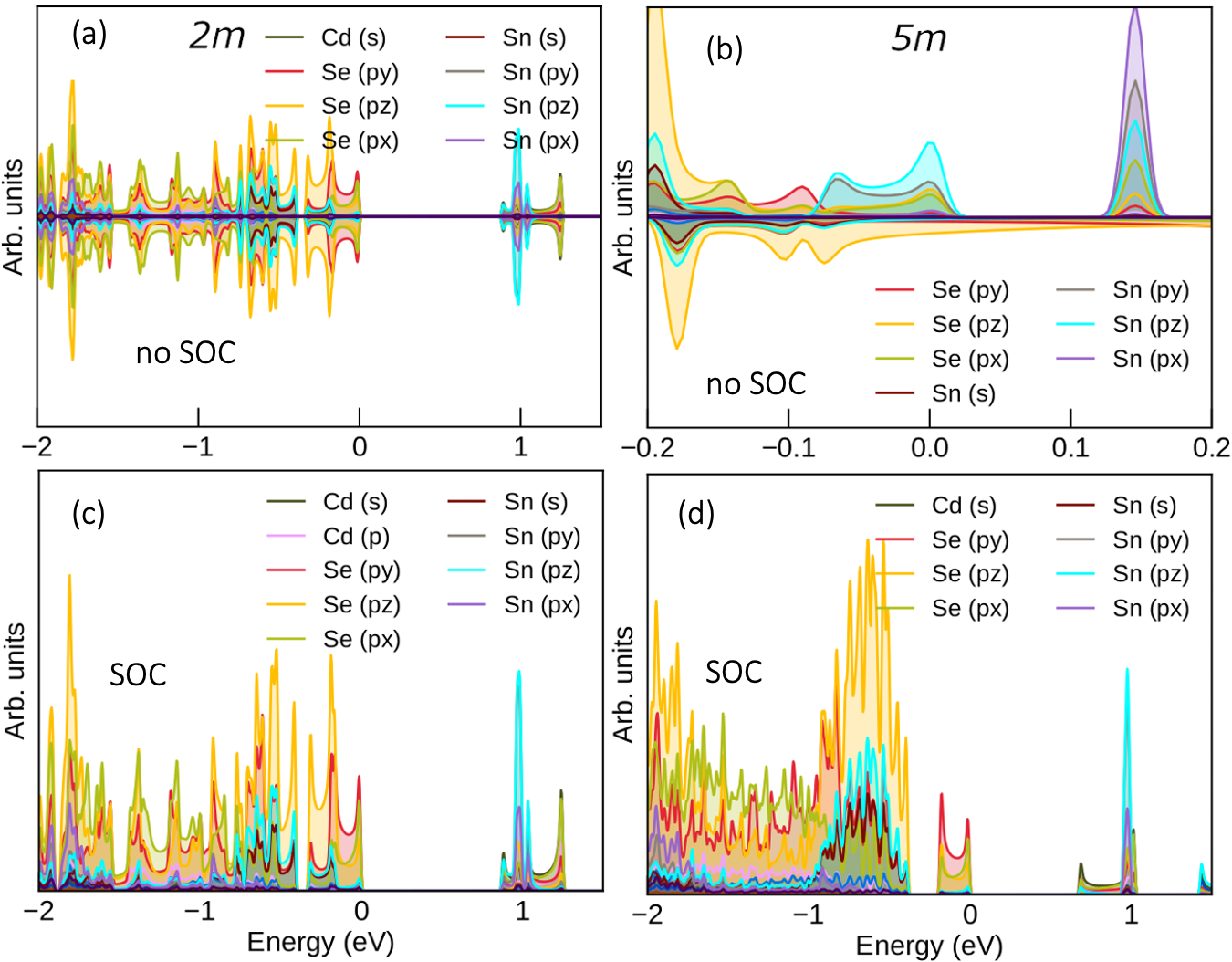}} 
\end{center}
\vskip -0.15in
{\bf Figure S5.} PDOS plots for CdSnSe armchair edge nanoribbons with the respective Fermi energy set to zero. Plots for width 2$m$ without SOC and with SOC are shown in the left top and left bottom panels whereas for width 5$m$ are shown in right top and right bottom panels. Upper panels show spin polarized PDOS.
\end{figure} 
\begin{figure}
\begin{center}
{\includegraphics[scale=0.5, valign=t]{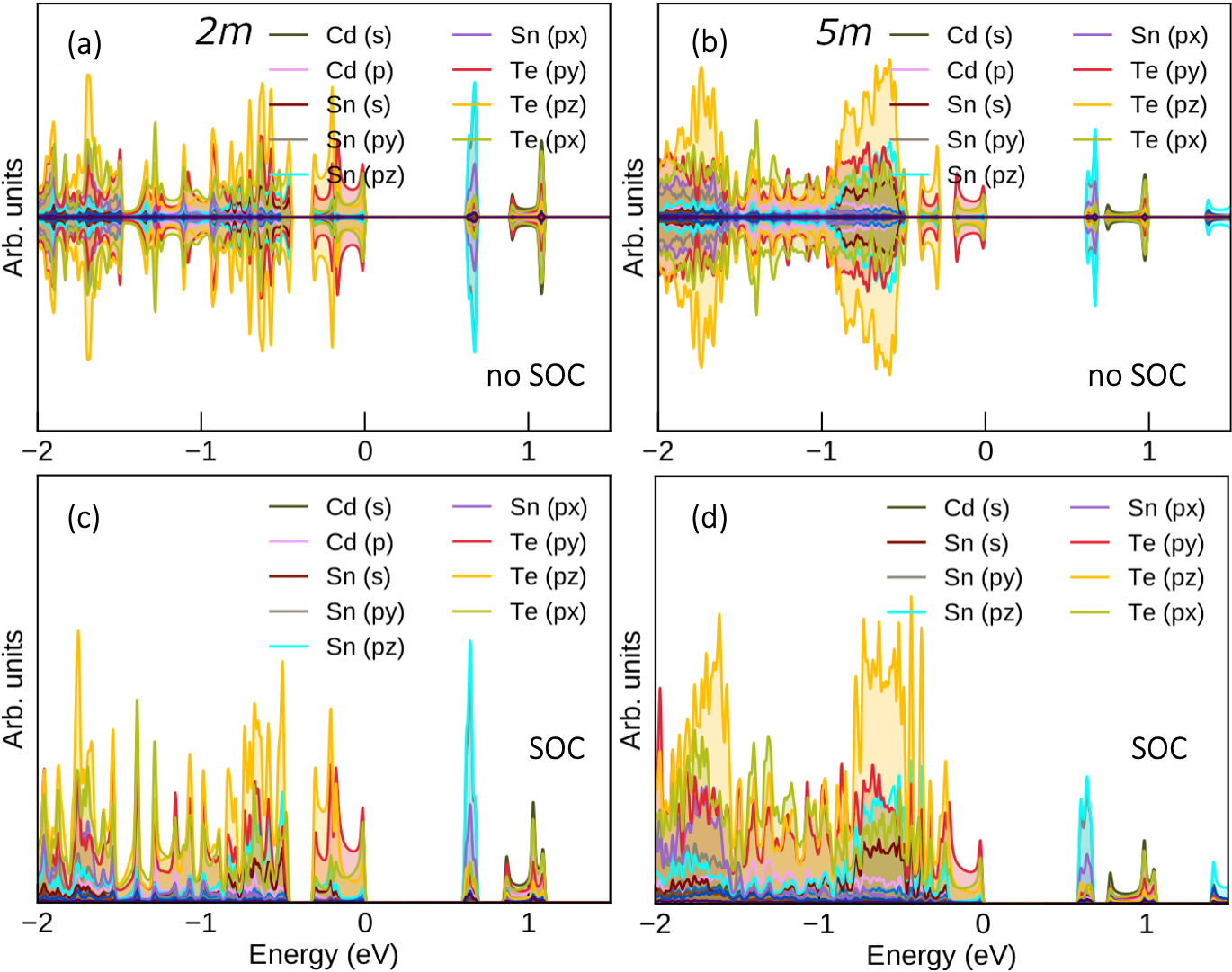}}
\end{center}
\hskip 0.2in 
{\bf Figure S6.} PDOS plots for CdSnTe armchair edge nanoribbons with the respective Fermi energy set to zero. Plots for width 2$m$ without SOC and with SOC are shown in the left top and left bottom panels whereas for width 5$m$ are shown in right top and right bottom panels. Upper panels show spin polarized PDOS.
\end{figure}